\documentclass[journal]{IEEEtran}
\usepackage{amsmath,amsfonts}
\usepackage{array}
\usepackage{cuted}
\usepackage{balance}
\usepackage{subcaption}
\usepackage{enumerate}
\usepackage{tabularx}
\usepackage{multirow}
\usepackage{xcolor}
\usepackage{textcomp}
\usepackage{stfloats}
\usepackage{flushend}
\usepackage[hyphens]{url}
\usepackage{tikz}
\usepackage{verbatim}
\usepackage{graphicx}
\usepackage[table]{xcolor}   
\def\BibTeX{{\rm B\kern-.05em{\sc i\kern-.025em b}\kern-.08em
    T\kern-.1667em\lower.7ex\hbox{E}\kern-.125emX}}
\usepackage{balance}
\usepackage[ruled,vlined,linesnumbered]{algorithm2e}
\usepackage{algpseudocode}
\DontPrintSemicolon
\SetKwInput{Input}{Input}
\SetKwInput{Output}{Output}
\SetKw{To}{ to }
\SetKw{Print}{Print }

\SetCommentSty{mycommfont}
\newcommand\saveAlgoCounter[1]{%
    \expandafter\xdef\csname saveAlgoCounter@#1\endcsname{\thealgocf}}
\newcommand{\repeatAlgoCaption}[2]{%
    \expandafter\let\expandafter\thealgocf\csname saveAlgoCounter@#1\endcsname
    \captionsetup{list=no}%
    \caption{#2 (continued from page~\pageref{#1})}
    \addtocounter{algocf}{-1}
}
\newcommand*{\affmark}[1][*]{\textsuperscript{#1}}
\newcommand*{\affaddr}[1]{#1}

\makeatletter
\def\@IEEEsectpunct{\ \,}
\makeatother

\begin{document}
\title{HyDRA: Deadline and Reuse-Aware Cacheability for Hardware Accelerators}
\author{{Ayushi Agarwal}\affmark[1], Anannya Mathur\affmark[1], and Preeti Ranjan Panda\affmark[1,2] \\
\affaddr{\affmark[1]\emph{Amar Nath and Shashi Khosla School of Information Technology, Indian Institute of Technology Delhi}, India}\\\affaddr{\affmark[2]\emph{Department of Computer Science and Engineering, Indian Institute of Technology Delhi}, India}
\thanks{This paper is an extended version of a manuscript accepted for publication in IEEE Transactions on Computer-Aided Design of Integrated Circuits and Systems (IEEE TCAD). It includes additional experiments, analyses, and evaluation details omitted from the published version due to page limitations. \par
A research grant from the R Systems Center of Excellence on Sustainable Artificial Intelligence supported this work.}
}

\markboth{}
{A. Agarwal, A. Mathur, P. R. Panda: HyDRA: Deadline and Reuse-Aware Cacheability for Accelerators}

\maketitle

\newcommand{\addhere}[1]{{\colorbox{yellow}{#1}}} 
\newcommand\fixthis{\protect{\colorbox{pink}{FIX THIS}}}
\newcommand\revisit{\protect{\colorbox{gray}{REDO}}}

\begin{abstract}
The system-level cache is a critical resource shared by processor cores and domain-specific accelerators in heterogeneous systems on chips (SoCs).
The strict QoS requirements of accelerators, such as deadlines, can lead to severe performance degradation of processor cores. Thus, managing the shared cache efficiently between cores and accelerators becomes crucial. State-of-the-art cache management techniques perform reuse-aware bypassing of accesses from cores with the help of reuse predictors to improve performance. However, architectural differences between accelerators and processor cores (often associated with deep cache hierarchies) can lead to significantly different reuse patterns at the shared cache. 
We propose a novel clustering-based methodology, \emph{LERN}, for learning and predicting the reuse behavior of hardware accelerators at the shared cache. 
We then propose a deadline and reuse-aware cache management strategy, \emph{HyDRA}, which explores a novel tradeoff between reuse and deadline awareness for performance efficiency. It uses LERN to dynamically predict the reuse behavior of the accelerator accesses and make bypass decisions to maximize the system throughput while meeting accelerator deadlines. 
We evaluate HyDRA across different workloads and varied accelerator configurations. It significantly improves the system performance and reduces the accelerator deadline miss rate.    
\end{abstract}

\begin{IEEEkeywords}
Shared Cache Management, Hardware Accelerators on MPSoCs, Accelerator Reuse Prediction.
\end{IEEEkeywords}

\section{Introduction}
\begin{figure}[htbp]
    \centering
    \includegraphics[width=\linewidth]{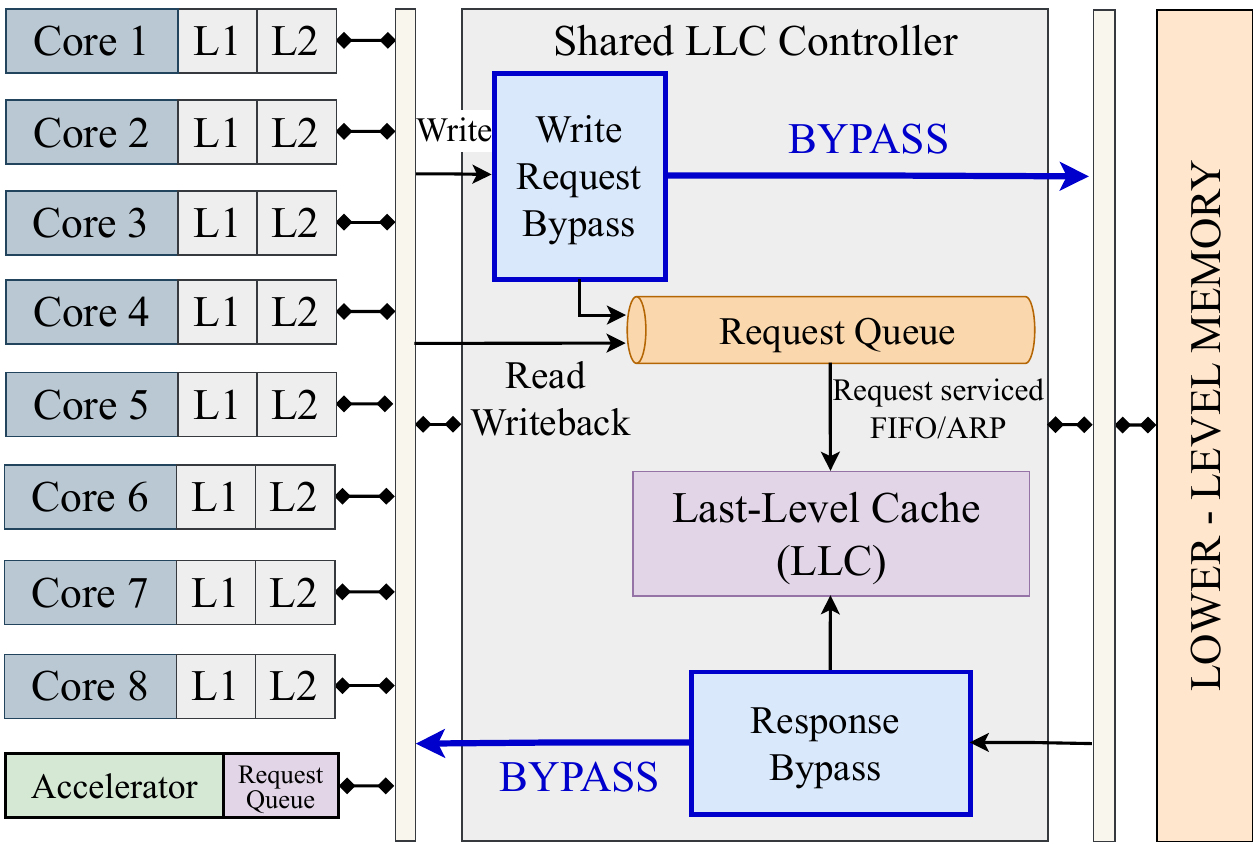}
    \caption{Overview of Heterogeneous System Architecture. We enhance \emph{gem5}~\cite{gem5} with integrated accelerators on the LLC~\cite{FLASH} 
    to add the write request/response bypass, and read response bypass for accelerator accesses (\emph{highlighted in \textcolor{blue}{blue}}).} 
    \label{fig:introduction}
\end{figure}

\IEEEPARstart{D}{omain}-specific hardware accelerators (HWA) have emerged as a compelling solution for meeting the increased computational demands of modern-day artificial intelligence (AI) applications. 
Recent trends in SoC architecture show that accelerators are integrated in the system to share the on-chip last-level cache (LLC) with the other processing elements, such as processor cores~\cite{AMDRyzen7040, QSnap888,  kirin9000}. Figure \ref{fig:introduction} provides an overview of such a heterogeneous system architecture. {The shared cache services read misses and writeback traffic from private caches of processor cores. Requests that miss in the shared cache are forwarded to off-chip memory, and returned data is written back to the shared cache. In systems with integrated accelerators, the shared cache must also handle read and write accesses from the accelerator.}
Hence, integrating accelerators on this critical shared resource increases its space and bandwidth contention. 
Therefore, intelligent management of the LLC space, bandwidth, or both, is crucial so that accelerators can meet their deadlines and maintain their quality of service (QoS) without degrading the performance of the rest of the system, including cores. 

The efficient management of shared cache space between processor cores~\cite{UCP, Intel-CAT, PACP2020, PRCP} and between CPUs and GPUs~\cite{CSCPU-GPU, tlpawarecachecpu-gpu} has been comprehensively studied in MPSoCs. Shared cache bandwidth management between cores ~\cite{feliu2017, REAL} and between cores and accelerators~\cite{FLASH} has also only recently garnered attention.
However, the interaction between cores and accelerators in the context of shared cache space remains unexplored. 
The most commonly used techniques for managing cache capacity across different computational units are static or dynamic cache space partitioning and selective caching or bypassing of memory accesses. Effective LLC bypassing depends heavily on the nature of the computational unit (CPU, GPU, or HWA) and on application characteristics, such as data locality and LLC reuse behavior. {Prior work has shown that locality-aware cache bypass can reduce contention in shared caches~\cite{OBM, SHIP}. As shown in Figure \ref{fig:introduction}, the cache controller may bypass certain accesses based on reuse predictions, such as accelerator write requests and read responses from memory for both cores and accelerators (shown in \emph{blue}), allowing them to bypass the shared cache and directly access main memory. }
The data locality of processor applications has been extensively studied to gain benefit from the on-chip cache hierarchies. 
Even though memory and data locality-aware design space optimizations~\cite{flexflow, eyerissAccel, cadosys} have been targeted for ML accelerators to optimize their performance, power, and area, their off-chip memory access patterns or data locality seen at the LLC have not been explored in CPU-HWA heterogeneous SoCs. 
Due to the architectural differences between processor cores, which are often associated with deep cache hierarchies, and accelerators, their reuse patterns observed at the LLC might differ significantly. 
Additionally, an efficient bypassing strategy, based on reuse behavior, designed for the processor applications might perform poorly with HWAs, negatively impacting the overall system performance. 

We motivate that the reuse analysis performed on processor applications does not necessarily hold for accelerators, and show that the state-of-the-art reuse predictors perform sub-optimally for accelerators. We propose a novel clustering-based methodology, \emph{LERN}, for learning accelerator reuse behavior at the LLC. 
We finally propose \emph{HyDRA}, a deadline- and reuse-aware policy for dynamically deciding the cacheability of memory accesses from accelerators and cores at the LLC depending on the accelerator's progress and reuse behavior. 
If the accelerator is in line to meet the deadline, HyDRA reallocates the cache space to processor cores by aggressively bypassing accelerator accesses from the LLC and positively impacting the performance of cores. 
The bypass contenders are determined by adding reuse awareness with offline-trained LERN predictors. 
If progress is insufficient to meet the deadline, accesses without reuse are bypassed. 
This brings out a novel and interesting tradeoff between deadline and reuse awareness. Without deadline awareness, a good reuse-aware policy would bypass accesses based solely on the reuse behavior, which might lead to deadline misses. A deadline-aware policy without good reuse awareness can help meet the deadline, but might not achieve better performance for cores. 
To our knowledge, this is the first comprehensive work to study the cache space contention and management between accelerators and cores. 
We make the following contributions:
\begin{enumerate}
    \item We introduce a crucial tradeoff between deadline and reuse awareness to solve a novel problem of dynamically managing the LLC space between accelerators and cores. 
    \item We propose a reuse pattern detection framework, LERN, which uses K-means clustering to learn the reuse-based similarities between the accelerator's memory accesses and partition them into clusters with similar reuse behavior. 
    LERN efficiently forms distinct clusters of memory accesses to predict their reuse behavior dynamically.
    \item We propose a novel approach, HyDRA, to improve the system performance by deadline and reuse-aware dynamic LLC bypass while meeting accelerator deadlines. The bypass aggressiveness for accelerator accesses depends on the progress made in meeting the deadline. 
    Deadline awareness conflicts with reuse awareness, as several accesses with low reuse properties might be cached to ensure meeting the deadlines, and vice versa. HyDRA achieves a balance between deadline and reuse awareness. 
    \item We enhance the \emph{gem5}~\cite{gem5} full-system simulator with accelerators integrated on the LLC~\cite{FLASH} to add accelerator write request/response bypass as well as read response bypass at the LLC.  We plan to integrate our infrastructure enhancements into the open-source gem5 repository. 
\end{enumerate}

We evaluate HyDRA across multiple accelerator architectures and machine learning (ML) models and show that it significantly improves cores' performance while reducing accelerator deadline misses. {The approach also extends to systems where the first shared level between cores and accelerators is an off-chip L4 cache rather than an on-chip LLC.}

\section{Related Work}
\label{sec:relwork}
\subsection{Shared Cache Management}
Shared cache space and/or bandwidth management have received considerable attention for homogeneous multi-processor SoCs and heterogeneous SoCs with GPUs. 
Shared cache space has been managed by techniques such as static or dynamic space partitioning~\cite{UCP,Intel-CAT, PACP2020, PRCP, tlpawarecachecpu-gpu}, efficient cache replacement~\cite{RRIP, LRFU}, and cache bypass~\cite{CSCPU-GPU, OBM, bfp, sdbp, IntelIA32bypass}. These techniques are driven by features such as cache utility~\cite{UCP, Intel-CAT}, prefetch awareness~\cite{PACP2020}, and data locality or reuse awareness~\cite{PRCP, OBM, bfp, RRIP, sdbp}. 

Shared cache bandwidth management between CPUs and accelerators is also crucial.
FLASH~\cite{FLASH} proposes a dynamic deadline-aware shared cache management policy to partition its bandwidth between cores and accelerators. Depending on the accelerator's progress, the shared cache bandwidth is reallocated to cores to improve IPC speedup. {However, it assumes that all accelerator requests should be serviced by the shared cache, which can lead to inefficient utilization in heterogeneous systems where accelerator accesses exhibit diverse reuse behavior.}
DPCP~\cite{DPCP}, on the other hand, jointly uses data prefetching and static cache partitioning to manage cache space between cores and accelerators.
However, DPCP evaluates a small system with a single CPU and a single accelerator and does not target QoS. Hence, the shared cache space management between CPUs and accelerators remains unexplored. 

\subsection{Reuse Analysis and Prediction: Uses and Methodologies}

{Researchers have extensively used reuse pattern analysis and prediction for optimizing web caching~\cite{learningforwardRD}, and on-chip caches by efficient cache replacement~\cite{SHIP, RRIP, LRFU, sdbp, MPPPB, perceptronLearnofRP, Glider, RLR, hawkeye, mockingjay, PACIPV2025, parrot, Stormbird, kpc, discrteCacheInsertion}, prefetching~\cite{LearningMAPrefetch}, and selective cache bypass \cite{OBM, bfp, sdbp, perceptronLearnofRP}.}
These predictions can broadly impact the cache performance by improving the cache hit rate and mitigating cache contention and interference. 

Heuristics-based solutions for reuse prediction, such as LRU, MRU, LRFU~\cite{LRFU}, and RRIP \cite{RRIP} have been proposed for efficient cache replacement. 
However, these solutions usually work well with commonly observed, limited cache access patterns despite low hardware overhead. 
More recently, learning-based algorithms \cite{OBM, SHIP, bfp, sdbp, MPPPB, perceptronLearnofRP, hawkeye, mockingjay, PACIPV2025, parrot} have been presented for learning cache access patterns based on the past accesses, using standard features such as the Program Counter (PC), memory address, and reference count to predict the reuse probability of a future memory access for efficient cache replacement or bypass. 
These predictors often compromise the farsightedness of the reuse patterns for hardware efficiency. 
Deep Learning and Reinforcement Learning algorithms \cite{learningforwardRD, Glider, RLR, LearningMAPrefetch} have also been explored for learning program behavior. {LSTM-driven learning of reuse behavior for memory addresses can incur significant hardware and training overhead. Since the size and training time of the LSTM grow in proportion to the number of inputs, previous work has shown that memory addresses are not feasible for LSTM training, as the number of unique memory addresses can be very large~\cite{Glider}. However, PCs are more suitable for training LSTMs. }
Glider \cite{Glider} proposed an online, lighter SVM model for efficient cache replacement, using features learned by an offline-trained LSTM model on an unordered PC trace to mitigate the high computational and hardware footprint of LSTM models, which are unsuitable for hardware caches. {RLR~\cite{RLR} and Stormbird~\cite{Stormbird} use reinforcement learning to learn an LLC replacement policy based on access, cache line, and set information.} These techniques are often challenging to converge and generalize.

Most of these techniques depend on processors' features, such as PC, and are efficient for applications that show consistent reuse patterns. 
{Hence, they are challenging to scale to accelerators that do not have such features, and their memory access patterns often exhibit long-range temporal correlations that exceed the short training horizons of hardware-optimized online schemes. Offline training, on the other hand, can capture global reuse behavior of memory accesses and learn long-range temporal correlations that online predictors usually miss. Offline training shifts the learning complexity from silicon to software.} 
We combine static profiling of the accelerators' off-chip memory accesses with dynamic on-chip cache management achieved by selective cache bypassing based on reuse awareness \emph{to maximize processor cores' performance while meeting accelerators' deadlines. }

\section{Motivation}
\label{sec:motivation}

\begin{figure}[t!]
    \centering
    \begin{subfigure}[b]{\linewidth}
        \centering
        \includegraphics[width=\linewidth]{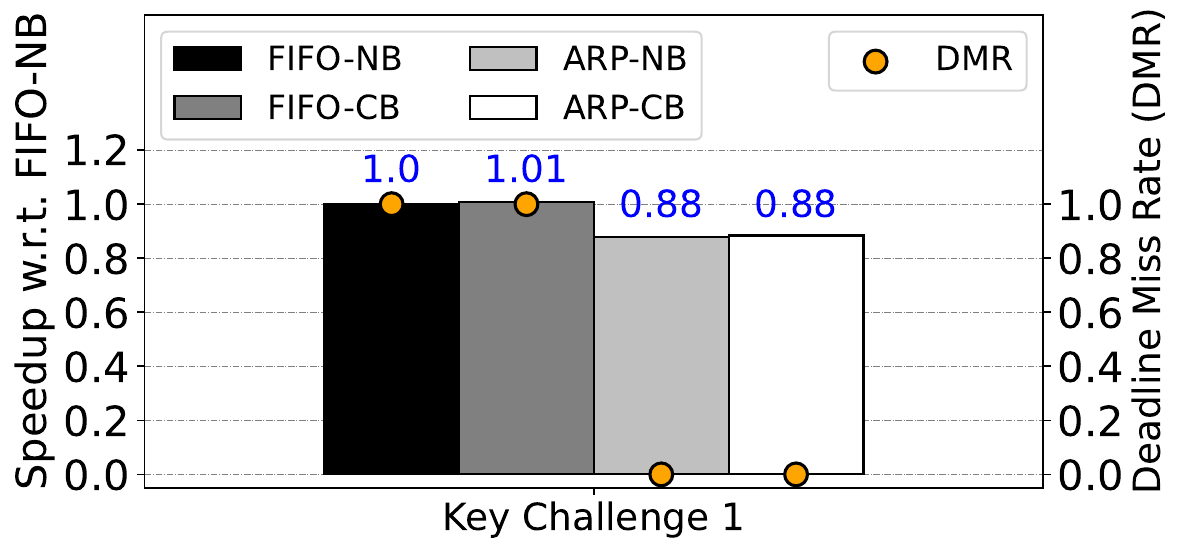}
        \caption{Accelerator-friendly Bandwidth Allocation with Core Bypass.}
        \label{fig:m1}
    \end{subfigure}
    \par
    \begin{subfigure}[b]{\linewidth}
        \centering
        \includegraphics[width=\linewidth]{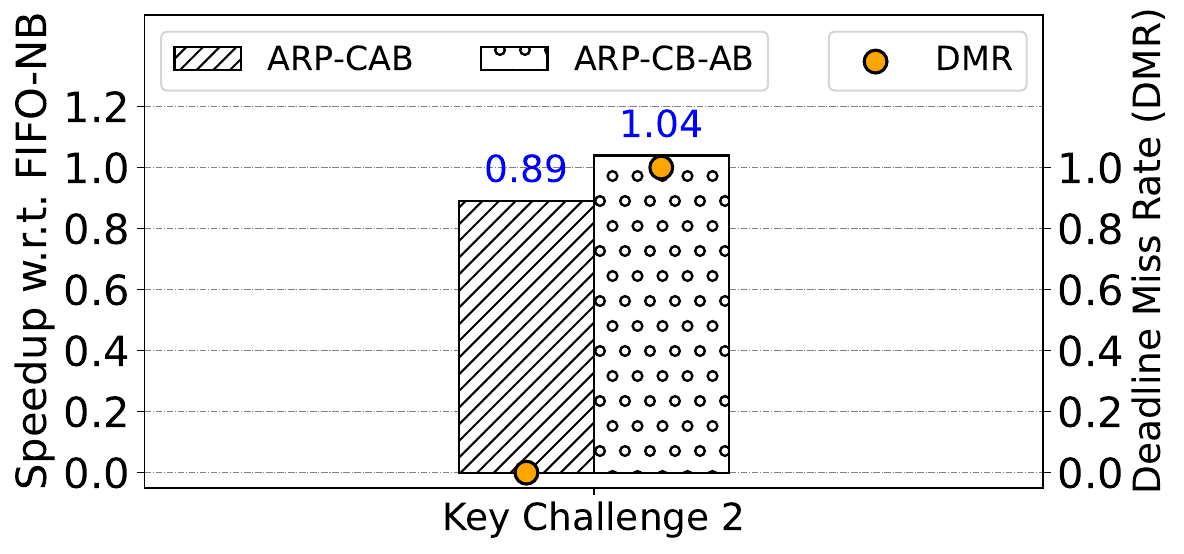}
        \caption{Performance with Shared vs. Private Reuse Predictors.}
        \label{fig:m2}
    \end{subfigure}
    \par
    \begin{subfigure}[b]{\linewidth}
        \centering
        \includegraphics[width=\linewidth]{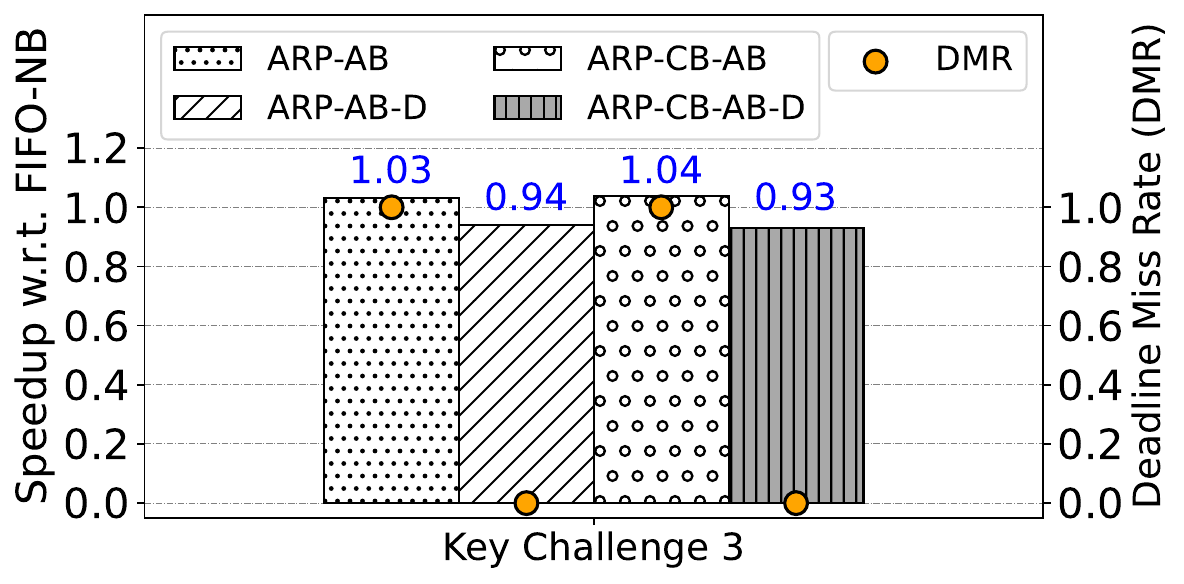}
        \caption{Deadline and Reuse-Aware Cacheability for Accelerators.}
        \label{fig:m3}
    \end{subfigure}
    \caption{Key Motivational Challenges 1, 2, and 3. Mean performance speedup and DMR with state-of-the-art LLC management policies.
    FIFO, ARP: LLC Arbitration Policies. {NB: \textbf{N}o \textbf{B}ypass, CB/AB: \textbf{C}ore/\textbf{A}ccelerator bypass with \textbf{B}aseline predictor, D: Deadline-aware.}} 
    \label{fig:motivation1}
\end{figure}

As shown in Figures \ref{fig:motivation1} and \ref{fig:motivation2}, we use three key challenges to motivate the need for deadline-aware, dynamic LLC management, with selective cache bypassing, between accelerators and cores, so that accelerator deadlines are met with minimal impact on cores' performance. 
{We use a state-of-the-art reuse prediction methodology (baseline or \emph{B}), such as SHIP~\cite{SHIP}, to identify the low-reuse bypass contenders from cores and accelerators. 
We use the following notation for representing the different cache management policies discussed ahead: \\
\emph{\{ $\!\mbox{Arbitration}\!-\!\mbox{C(Policy)}\!-\!\mbox{A(Policy)}\!-\!\mbox{Deadline}\!$ \}}, \\
where \emph{$\mbox{Arbitration}$} denotes the cache arbitration policy, \emph{$\mbox{C(Policy)}$} and \emph{$\mbox{A(Policy)}$} denote the reuse-aware bypass policies for core accesses and accelerator accesses at the cache, respectively, and \emph{$\mbox{Deadline}$} denotes deadline awareness.}

For processor cores, we monitor the total achieved throughput. For accelerators, we use a metric called Deadline Miss Rate (DMR), the ratio of the number of input sets for which the deadline is missed to the total number of input sets executed. We analyze the mean performance obtained on a system with eight processor cores running two different combinations of SPEC applications (1: \emph{omnetpp} (high-reuse) on all cores, 2: \emph{omnetpp} on four cores, \emph{mcf} (low-reuse) on four cores) and one TPU~\cite{GoogleTPU}-grade accelerator running the Tiny-YOLO network with configuration Config-1, as detailed in Tables \ref{tab:systemconfig} and \ref{tab:hwaconfig}. 
The deadline used for the accelerator is derived from a 10 inputs per second (IPS) image frame rate. In this case, an input is an image frame. 

\subsection{Key Challenge 1: Accelerator-friendly Bandwidth Allocation along with Core Bypass.}
\label{sec:keyInsight1}

\subsubsection{With insufficient bandwidth allocation to the accelerator, is reuse-aware core bypass sufficient to meet its deadline?}
\label{sec:keyInsight1a}

Without intervention, the LLC controller schedules the requests from accelerators and cores in {first-in, first-out (FIFO)} order with no bypass ({FIFO-NB}). 
FIFO arbitration provides LLC bandwidth on demand, and when contending with multiple processors, the accelerator might miss its deadline due to insufficient bandwidth (Figure \ref{fig:m1}). 
{Since accelerators also contend for cache space, this raises the question whether alleviating space contention with {reuse-aware cache bypass of the core accesses or \textbf{C}ore bypass} using the \textbf{B}aseline predictor (CB) is enough to meet the accelerator's deadline (FIFO-CB).}
With 2\% core accesses bypassed, depending on the application, and a marginal throughput increase of 1\%, the relieved cache space is insufficient to meet the deadline without sufficient bandwidth. 
Adding \emph{reuse-aware accelerator bypass} would worsen its execution time and not help meet the deadline. 
Therefore, LLC intervention is required first to provide the accelerator with sufficient bandwidth. 

\subsubsection{With accelerator-friendly bandwidth allocation, is reuse-aware core bypass sufficient to restore the lost performance of processor cores?}
\label{sec:keyInsight1b}

We use a static arbitration policy that prioritizes accelerator requests at each LLC access, called Accelerator Request Priority (ARP)~\cite{FLASH} with no bypass {(ARP-NB)}. Figure \ref{fig:m1} shows that ARP-NB meets the given deadline but with a significant degradation of up to 12\% in the cores' performance as static arbitration can lead to over-allocation of bandwidth to the accelerator. 
{To efficiently manage cache space with a static bandwidth allocation, we add reuse-aware \textbf{C}ore bypass with the \textbf{B}aseline predictor to see if it can restore the lost performance of cores while meeting the deadline (ARP-CB).} The accelerator meets the deadline with sufficient bandwidth, but selective core bypass alone cannot restore the cores' performance (Fig. \ref{fig:m1}).

\subsection{Key Challenge 2: Given sufficient bandwidth, can reuse-aware accelerator bypass, using shared reuse predictors with cores, restore the system performance?}
\label{sec:keyInsight2}

Dynamic reuse prediction methodologies for efficient replacement and bypass for cores have often shown that per-core private predictor tables perform comparably with shared but bigger predictor tables~\cite{SHIP}. 
Given a heterogeneous system with an accelerator, we explore whether the same hypothesis holds.  
{We compare two architectural configurations that are deadline-agnostic (denoted by the absence of \emph{$-\mbox{D}$}), use ARP arbitration, perform core and accelerator bypass based solely on reuse behavior: (1) with a shared baseline reuse predictor (ARP-CAB), and (2) with separate baseline reuse predictors (ARP-CB-AB). }
{As shown in Figure \ref{fig:m2}, with shared predictors, we observe minimal performance improvement due to a marginal increase in the mean core bypass rate (Core BR), and the accelerator bypass rate (Accel BR) closely follows the cores, signifying access pattern interference.}
With separate predictors, we observe a speedup of 4\% over FIFO-NB, and the accelerator shows a higher bypass rate that is not impacted by the interference from cores. We conclude that due to different access patterns, cores and accelerators should not share a reuse predictor.

\subsection{Key Challenge 3: Deadline and Reuse-Aware Cacheability for Hardware Accelerators.}
\label{sec:keyInsight3}

\subsubsection{Adding deadline awareness to reuse-aware bypass.}
\label{sec:deadlineAwareness}

In Section \ref{sec:keyInsight2}, we observed that reuse-aware accelerator bypass leads to deadline misses without deadline awareness. 
We add deadline awareness to dynamically adjust the bypass aggressiveness based on the accelerator's progress, which is monitored periodically after each epoch (of duration ET). 
{For meeting the required inputs per second rate (IPS), the ${M}$ memory accesses in one input set, determined by static workload profiling based on accelerator characteristics, should complete in time ${D}_{\mbox{\scriptsize sec}}$ (converted from IPS). 
These are communicated to the LLC controller at the beginning of the application’s execution.}
{To add \textbf{D}eadline awareness to the reuse-aware \textbf{C}ore and \textbf{A}ccelerator bypass using different \textbf{B}aseline predictors (ARP-CB-AB-D), the accelerator bypass begins when the number of accesses completed within the epoch exceeds the number of accesses required to be completed in the epoch according to its deadline ($M\times \mbox{ET}/D_{sec}$).}
From Figure \ref{fig:m3}, we observe that adding deadline awareness to reuse-aware bypass meets the deadline, but accelerator bypass rate drops significantly to less than 10\% from around 60\% (with ARP-CB-AB), and the cores' performance reduces.
This depicts a strong tradeoff between deadline and reuse awareness in bypassing decisions while optimizing performance with a deadline constraint. 

\subsubsection{Is the reuse predictor, designed for general-purpose applications, the right choice for hardware accelerators?}
\label{sec:diffAccelPredictor}

Figure \ref{fig:motivation1} shows that performing reuse-aware accelerator bypass with and without deadline awareness can lead to significant performance variation. Without deadline awareness, we observe a high DMR but a higher IPC speedup. With deadline awareness, the deadline is met, but the IPC drops significantly because of a lower accelerator bypass rate. This steep drop in the bypass rate raises an important question: Is the baseline reuse prediction methodology, designed for cores, the right choice for detecting reuse patterns from the accelerator?

\begin{figure}[htbp]
    \centering
    \includegraphics[width=\linewidth]{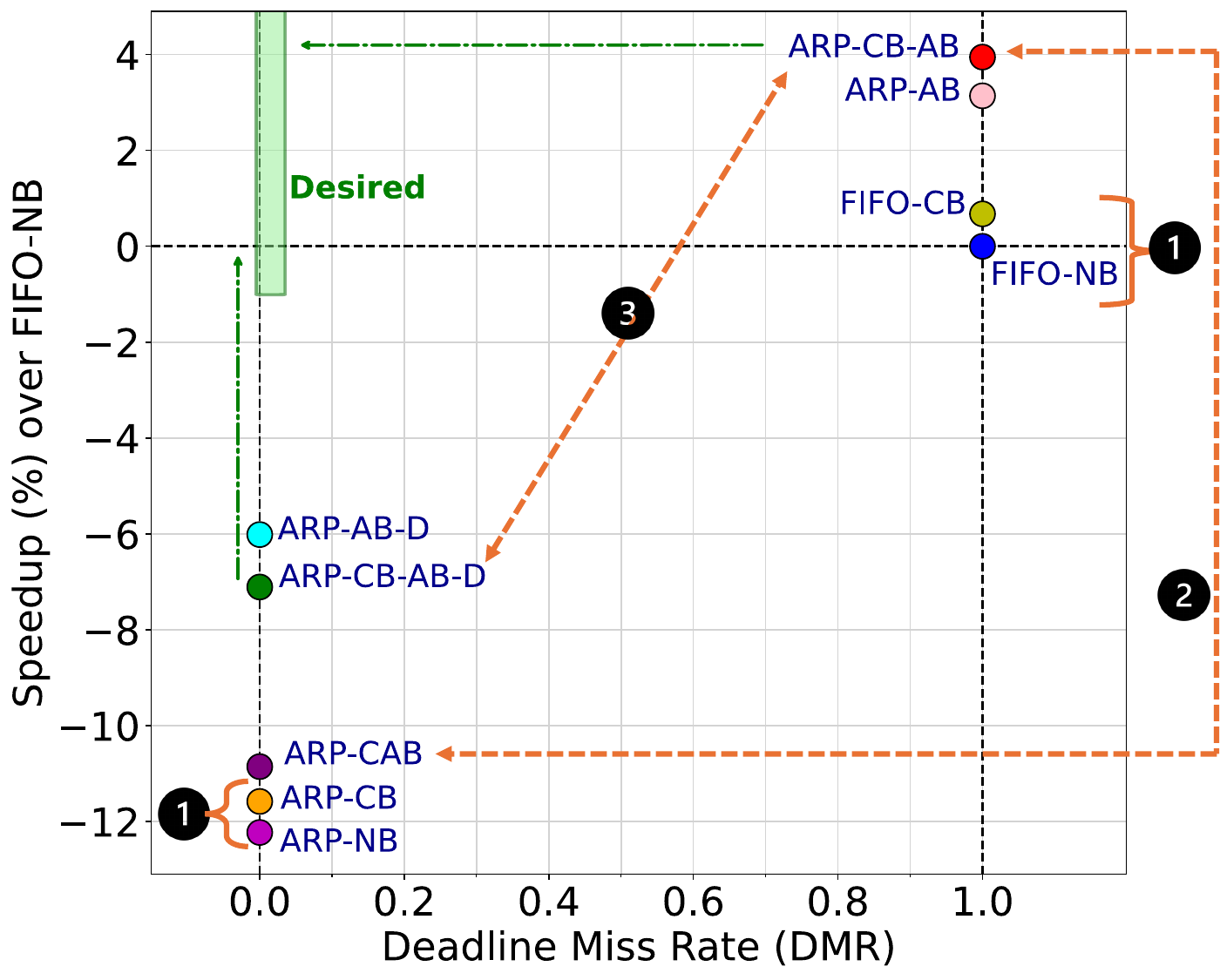}
    \caption{Limitations of the state-of-the-art cache management solutions and Key Motivational Challenges \protect\tikz[baseline=(X.base)] \protect\node (X) [draw, circle, fill=black, text=white, inner sep=1.5pt] {\small{1}};, \protect\tikz[baseline=(X.base)] \protect\node (X) [draw, circle, fill=black, text=white, inner sep=1.5pt] {\small{2}};, and \protect\tikz[baseline=(X.base)] \protect\node (X) [draw, circle, fill=black, text=white, inner sep=1.5pt] {\small{3}};.
    The desired region shown in green represents the potential solutions for optimizing the system throughput with deadlines.}
    \label{fig:motivation2}
\end{figure}

Figure \ref{fig:motivation2} highlights the three key challenges discussed earlier to emphasize the need for a shared cache management policy that addresses the tradeoff between reuse and deadline awareness and maximizes the system throughput while meeting the accelerator's deadline. The $y$-axis of the figure captures the performance speedup over the FIFO-NB baseline, and the $x$-axis denotes the accelerator's DMR. 
It shows that adding deadline awareness to reuse-aware bypass helps to meet the deadline. However, performance can be further improved with better reuse prediction. This points to the requirement of a different reuse prediction methodology for the accelerator rather than using the baseline predictors. 
The figure highlights the desired region (shaded in green) of optimization where a near-zero deadline miss rate is achieved with high system throughput.

\section{LERN: Clustering-based LEaRNing and Prediction of Reuse for Hardware Accelerators. }
\label{sec:lern}

\begin{figure}[htbp]
    \centering
    \includegraphics[width=\linewidth]{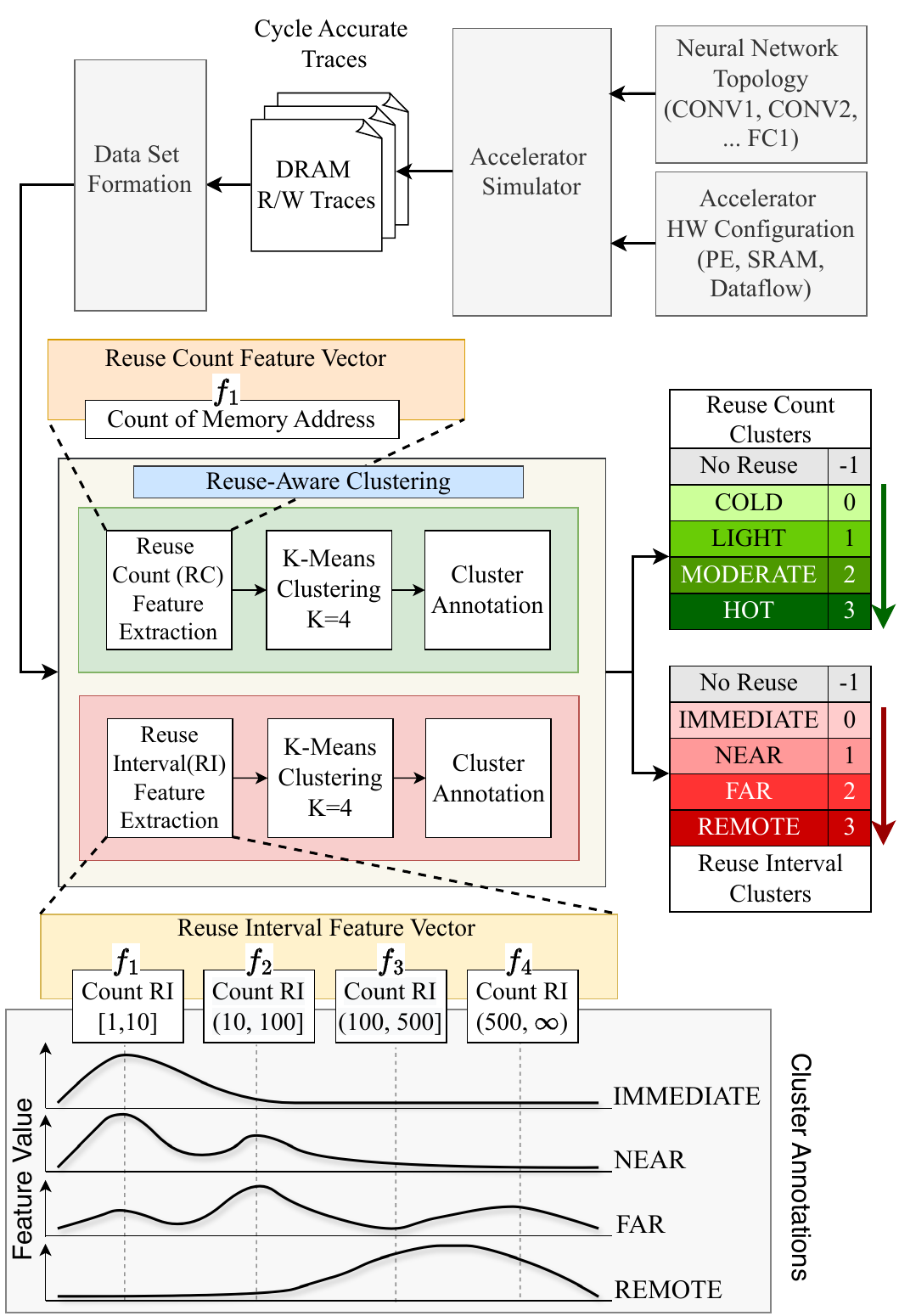}
    \caption{LERN Methodology Overview: Reuse Count and Reuse Interval based Clustering of Accelerator Memory Accesses. }
    \label{fig:lern}
\end{figure}

The reuse behavior of hardware accelerators can be characterized by their Reuse Interval (RI) and Reuse Count (RC) in the context of a memory access sequence. RI is defined as the number of accesses between two consecutive accesses to the same datum~\cite{learningforwardRD}. 
RC is defined as the number of times a single datum appears in the complete memory access sequence of the accelerator. 
Table \ref{tab:RIRCCalc} shows our interpretation of the RI and RC for a sample memory access sequence. 
We propose a statistical clustering-based learning methodology, called LERN (Figure \ref{fig:lern}), to statically profile the off-chip memory access sequence from accelerator simulators, such as SCALE-Sim~\cite{samajdar2020systematic}, and learn the reuse behavior quantified by RI and RC.
Using these two features, the complete memory access sequence is segmented into different reuse-based clusters, which are then used for LLC optimizations. 

\subsection{Dataset Formation}
\label{sec:datasetformation}
A memory access sequence is a series of ${M}$ memory accesses with single or multiple occurrences of ${N}$ unique addresses $a_n$, where $1 \le n \le N$.
To learn the reuse behavior at the shared cache, we generate a Reuse Signature Trace $\mbox{\emph{TR}}$ for all the unique cache addresses ($N_c$) in the original trace. Multiple addresses $a_n$ can map to a single cache line address $c_i$, where $1 \le i \le N_c \le N$. Therefore, the RI and RC of $c_i$ may differ from those of the addresses that map to it. In the example shown in Table \ref{tab:RIRCCalc}, ${M}=8$, ${N}=4$, and ${N}_c=2$.
The access sequence for cache line address $c_i$ is given by $\mbox{r}_i$, the sequence of memory access numbers (m) corresponding to the $c_i$ accesses, the sequence length being $T_i$, and $\mbox{r}_{i,j}$ being the $j^{th}$ element of sequence $\mbox{r}_i$.
For example, the cache line address $c_1$, as shown in Table \ref{tab:RIRCCalc}, occurs in the trace at $\mbox{m}\!=\!1, 2, 3, 6, 7$, hence, $\mbox{r}_1=\{1,2,3,6,7\}$.

\begin{table}[htbp]
  \caption{Reuse Interval, Count for a Memory Access Sequence. Memory addresses $\{a_1, a_2\} \in \mbox{cache line } c_1$, and $\{a_3, a_4\} \in \mbox{cache line } c_2$.}
  \label{tab:RIRCCalc}
  \begin{tabularx}{\linewidth}{|>{\raggedright\arraybackslash}p{2.9cm}|>{\centering\arraybackslash}p{0.252cm}|>{\centering\arraybackslash}p{0.252cm}|>{\centering\arraybackslash}p{0.252cm}|>{\centering\arraybackslash}p{0.252cm}|>{\centering\arraybackslash}p{0.252cm}|>{\centering\arraybackslash}p{0.252cm}|>{\centering\arraybackslash}p{0.252cm}|>{\centering\arraybackslash}p{0.252cm}|}
    \hline
    \rowcolor{gray!20} \textbf{Memory Access Number (m)} & \multirow{2}{*}{1} & \multirow{2}{*}{2} & \multirow{2}{*}{3} & \multirow{2}{*}{4} & \multirow{2}{*}{5} & \multirow{2}{*}{6} & \multirow{2}{*}{7} & \multirow{2}{*}{8} \\
    \hline
    \textbf{Memory Address} & $a_1$ & $a_2$  & $a_1$  & $a_3$  & $a_4$ & $a_1$ & $a_2$ & $a_3$ \\
    \hline
    \rowcolor{gray!20} \textbf{Reuse Interval (RI)} & 2 & 5 & 3 & 4 & -1 & -1 & -1 & -1 \\
    \hline
    \textbf{Reuse Count (RC)} & 1 & 1 & 2 & 1 & 1 & 3 & 2 & 2 \\
    \hline
    \hline
    \rowcolor{gray!20} \textbf{Cache line Address} & $c_1$ & $c_1$  & $c_1$  & $c_2$  & $c_2$ & $c_1$ & $c_1$ & $c_2$ \\
    \hline
    \textbf{Reuse Interval (RI)} & 1 & 1 & 3 & 1 & 3 & 1 & -1 & -1 \\
    \hline
    \rowcolor{gray!20} \textbf{Reuse Count (RC)} & 1 & 2 & 3 & 1 & 2 & 4 & 5 & 3 \\
    \hline    
  \end{tabularx}
\end{table}

The Reuse Signature Trace \emph{TR} is a sequence of Reuse Vectors (RV) for all the unique cache line addresses. 
For each address ${c}_i$, $\mbox{RV}_i$ is a sequence that records the RI for all its $T_i$ occurrences.
Each RI of address $c_i$ at occurrence $j$ or $\mbox{RI}_{i,j}$ is defined as $\mbox{r}_{i,j+1} - \mbox{r}_{i,j}$ with the last interval being -1.

\subsection{RI and RC Feature Extraction}
\label{sec:featureVector}

We use the Reuse Signature Trace $\mbox{\emph{TR}}$ to extract the RI and RC features, $F_{RI}$ and $F_{RC}$, respectively.
$F_{RC}$ is a sequence of features $\{T_1, T_2, .., T_{N_c} \}$, where $T_i$ (sequence length of $\mbox{RV}_i$) is the reuse count of the unique cache line address $c_i$ occurring in the memory access sequence.
$F_{RI}$ is a sequence of feature vectors for all cache line addresses. Each feature vector is a tuple of four features, $f_1, f_2, f_3, f_4$, representing the frequency of the RI being in different bins, such as $[1,10]$, $(10, 100]$, $(100, 500]$, and $(500,\infty)$, respectively. 
{These bins are chosen based on system cache associativity (in this case, 16 ways) and interference characteristics, where RIs below 10 likely survive in the cache, intermediate intervals may or may not survive, and larger intervals represent long-range reuse with higher eviction probability.}
For the example in Table \ref{tab:RIRCCalc}, $F_{RC} = \{5, 3\}$ and $F_{RI} = \{\{4,0,0,0\}, \{2,0,0,0\}\}$ for $N_c=2$.

\subsection{K-Means Clustering and Cluster Annotations}
\label{sec:clustering}

We formulate the reuse prediction problem as a pattern-detection and classification problem to classify memory accesses based on reuse. 
{As shown in Figure \ref{fig:lern}, we use K-Means clustering on the RI and RC features ($F_{RI}$, $F_{RC}$) to classify memory accesses into four clusters (K-Means input $n=4$) with predefined reuse behavior. Four clusters are used to provide finer-grained reuse classification than the binary live/dead decisions commonly used in LLC predictors. Binary classification collapses all reuse behavior into two decisions, and is insufficient for deadline-aware bypass because accelerator accesses exhibit diverse reuse distances, and bypass aggressiveness must adapt to different levels of the accelerator's progress.}

\subsubsection{K-Means Clustering - Reuse Count.}
The K-means clustering algorithm uses the RC features ($F_{RC}$) to cluster the memory accesses with similar reuse counts into four clusters of type ``Cold" (0), ``Light" (1), ``Moderate" (2), and ``Hot"  (3). {These clusters represent increasing levels of reuse observed in the reuse signature trace, where Cold corresponds to the smallest reuse counts, and Hot corresponds to accesses with the highest reuse counts.}

\subsubsection{K-Means Clustering - Reuse Interval.}
The K-Means clustering algorithm uses the normalized RI features ($F_{RI}$) to cluster the memory accesses with similar RI distribution into four categories of type ``Immediate" (0), ``Near" (1), ``Far" (2), and ``Remote" (3). 
The K-Means algorithm categorizes the addresses based on which feature or group of features is dominant in the feature vectors.  
Post-clustering, we use the relative values of the features in the de-normalized cluster centers to annotate the clusters into the four reuse categories, as illustrated in Figure \ref{fig:lern}. 

\begin{enumerate}
    \item ``Immediate" RI Cluster: The addresses in the cluster with dominant $f_1$ are marked as Immediate RI. 
    \item ``Near" RI Cluster: With multiple dominant features, if $f_1$ holds higher relative dominance over the other features, such as $f_2$, that cluster is annotated as Near RI. Most address occurrences show low RI, although some can show higher RI. 
    \item ``Far" RI Cluster: When $f_2$ holds dominance over $f_1$ or $f_3$ and $f_4$ holds relative dominance over $f_1$ but not $f_2$, the cluster is annotated as Far RI. 
    Most address occurrences show RI in the range represented by $f_2$.
    \item ``Remote" RI Cluster: Clusters with dominant $f_3$ or $f_4$ are annotated as Remote RI. Most occurrences show high RI, although some occurrences can show low RI. 
\end{enumerate}

The memory accesses with only one occurrence in the trace are classified into the ``No Reuse (-1)" RC and RI cluster. 

\begin{figure}[htbp]
    \centering
    \begin{subfigure}[b]{\linewidth}
        \includegraphics[width=\linewidth]{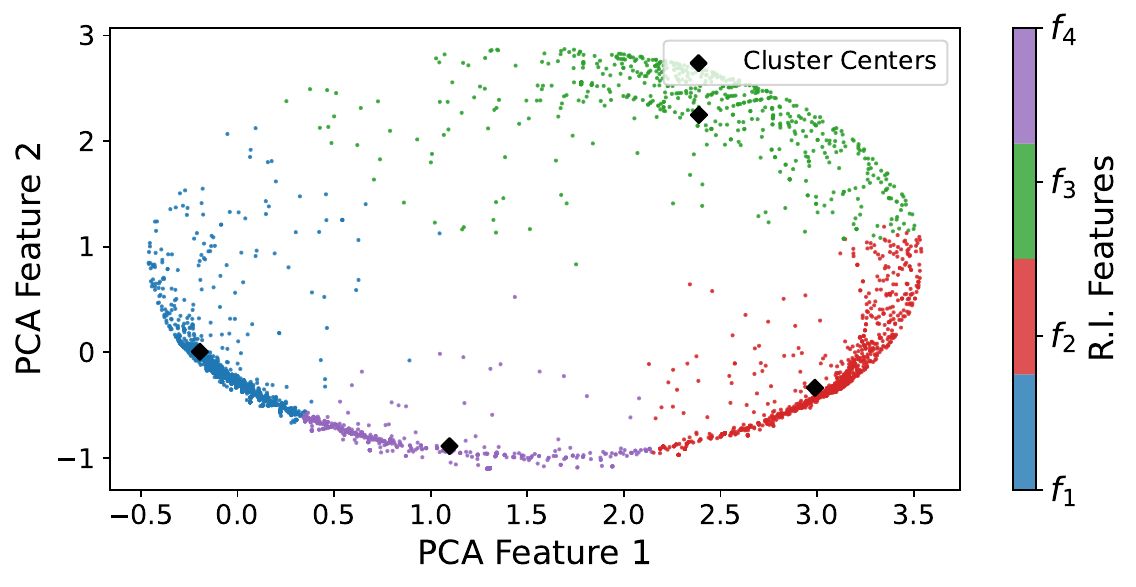}
        \caption{Tiny-YOLO Conv-1 layer. Clustering silhouette score ($>$ 90\%). Good Feature Separation $\rightarrow$ Efficient clustering in 4 clusters.}
        \label{fig:rd_features_c1}
    \end{subfigure}
    \par
    \begin{subfigure}[b]{\linewidth}
        \includegraphics[width=\linewidth]{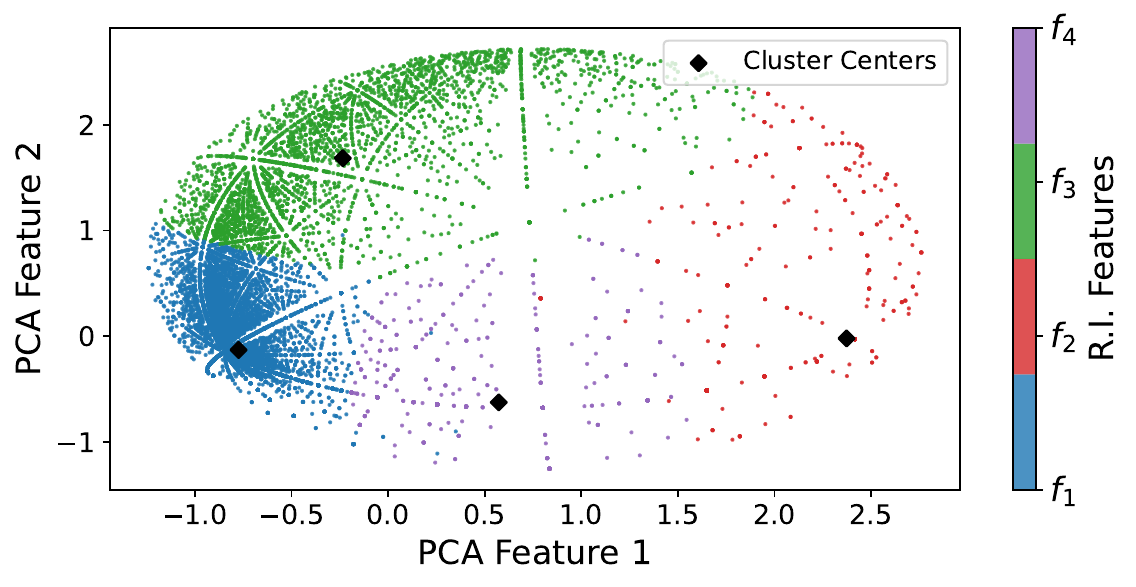}
        \caption{Tiny-YOLO Conv-7 layer. Clustering silhouette score $>$ 75\%. Satisfactory Feature Separation for clustering within 4 clusters.}
        \label{fig:rd_features_c7}
    \end{subfigure}
    \caption{PCA projection in 2-D for the 4-D RI features $(f_1, f_2, f_3, f_4)$ to show the achieved Feature Separability on Accelerator Configuration: Config-3.}
    \label{fig:rd_features}
\end{figure}

\subsection{Clustering Performance and Prediction Accuracy.}
\label{sec:clusterPerf}

To determine the clustering performance, we observe the separability of the 4-D RI features ($F_{RI}$) between the clusters in a 2-D space with PCA dimensionality reduction (Figure \ref{fig:rd_features}), along with the silhouette score achieved by K-Means on layers Conv1 and Conv7 of the accelerator configuration Config-3 (Table \ref{tab:hwaconfig}). 
K-Means achieves good feature separability with a high silhouette score representing high clustering accuracy within four clusters on Conv-1. 
On Conv-7, the 75\% silhouette score indicates a need for more clusters. {However, increasing the number of clusters raises hardware overhead, while the marginal improvement in reuse discrimination diminishes at the LLC relative to the hardware cost. 
Additional clusters may be justified only in highly constrained systems with strict QoS or tighter deadlines, where small improvements in bypass decisions can yield measurable performance gains for cores.}

\begin{figure}[htbp]
    \centering
    \begin{subfigure}[b]{\linewidth}
        \centering
        \includegraphics[width=\linewidth]{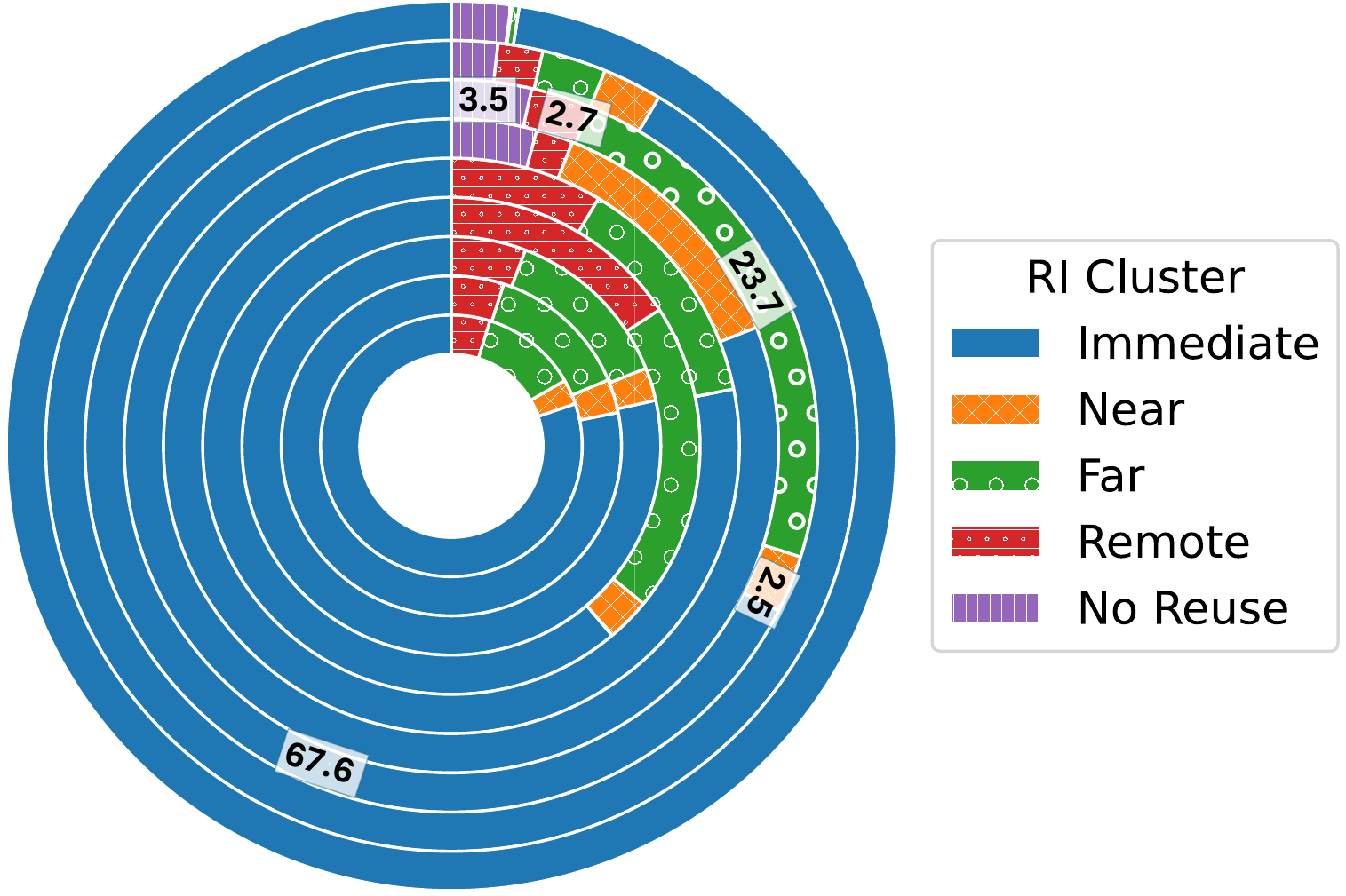}
        \caption{Memory Access Distribution from RI-based Clustering.}
        \label{fig:RI-cluster}
    \end{subfigure}
    \par
    \begin{subfigure}[b]{\linewidth}
        \centering
        \includegraphics[width=\linewidth]{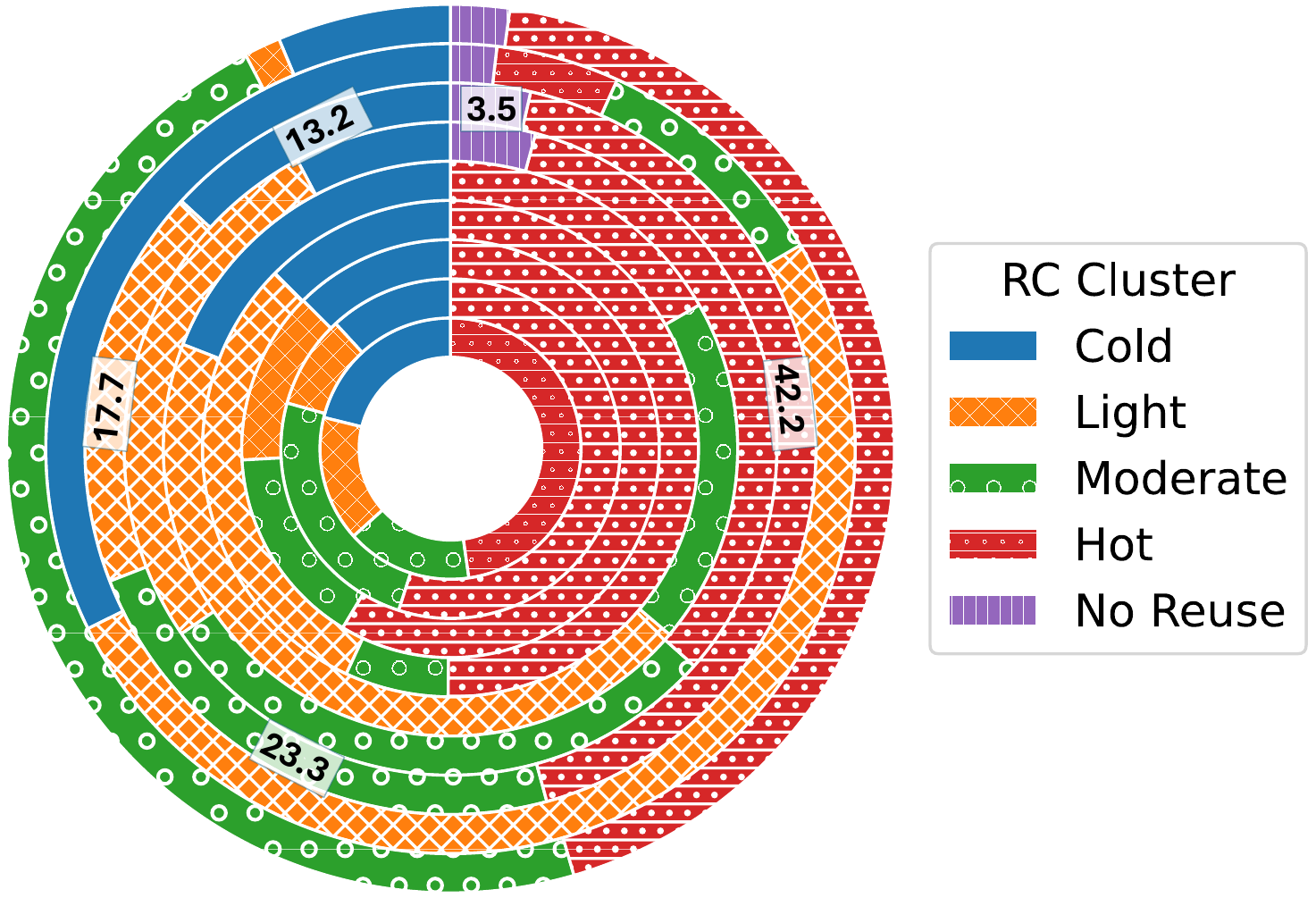}
        \caption{Memory Access Distribution from RC-based Clustering}
        \label{fig:RC-cluster}
    \end{subfigure}

    \caption{Percentage of memory accesses clustered in different RI and RC clusters for different layers of Tiny-YOLO on Accelerator Configuration: Config-3; The order of layers follows outward concentric circles.}
    \label{fig:Perc-mem-access}
\end{figure}

Figure \ref{fig:Perc-mem-access} shows the percentage distribution of memory accesses in all the layers of accelerator Config-3 across the different RI and RC clusters.
{Figure \ref{fig:RI-cluster} shows that more than 60\% of the memory accesses in all the layers are classified in the Immediate RI cluster, indicating strong data locality, as the simulated accelerator has small on-chip SRAMs, thereby showing high sensitivity to accelerator hardware.}
A significant fraction of the accesses will be predicted to have good reuse if only RI is used to predict reuse behavior. 
Figure \ref{fig:RC-cluster} shows a more uniform distribution of memory accesses across all the RC clusters, depicting varied reuse count behavior. 
Hence, considering RI and RC provides a more comprehensive view of the reuse behavior and could lead to better LLC optimizations. 

{LERN captures the most probable reuse behavior for a memory access instead of learning the exact temporal variation of its reuse intervals throughout the execution.}
To calculate the mean accuracy of the LERN RI-based prediction, we compare the cluster allocations with the actual RI of the memory accesses at different occurrences in the memory access sequence.
For an ``Immediate" RI cluster access $a_i$ with $\mbox{RV}_i\!=\!\{5, 20, 9, 6\}$, the accuracy is calculated with three correct and one incorrect RI prediction. 
LERN achieves 100\%, 100\%, 87.4\%, 89.8\%, 95.6\%, 93.9\%, 93.2\%, 93.6\%, 99.7\%, and 94.6\% accuracy on the accelerator configurations Config-1 to Config-10 (Table \ref{tab:hwaconfig}), respectively.

\section{HyDRA: Deadline and Reuse-Aware Cacheability for Hardware Accelerators}
\label{sec:hydra}

\begin{figure}[htbp]
    \centering
    \begin{subfigure}[b]{\linewidth}
        \centering
        \includegraphics[width=\linewidth]{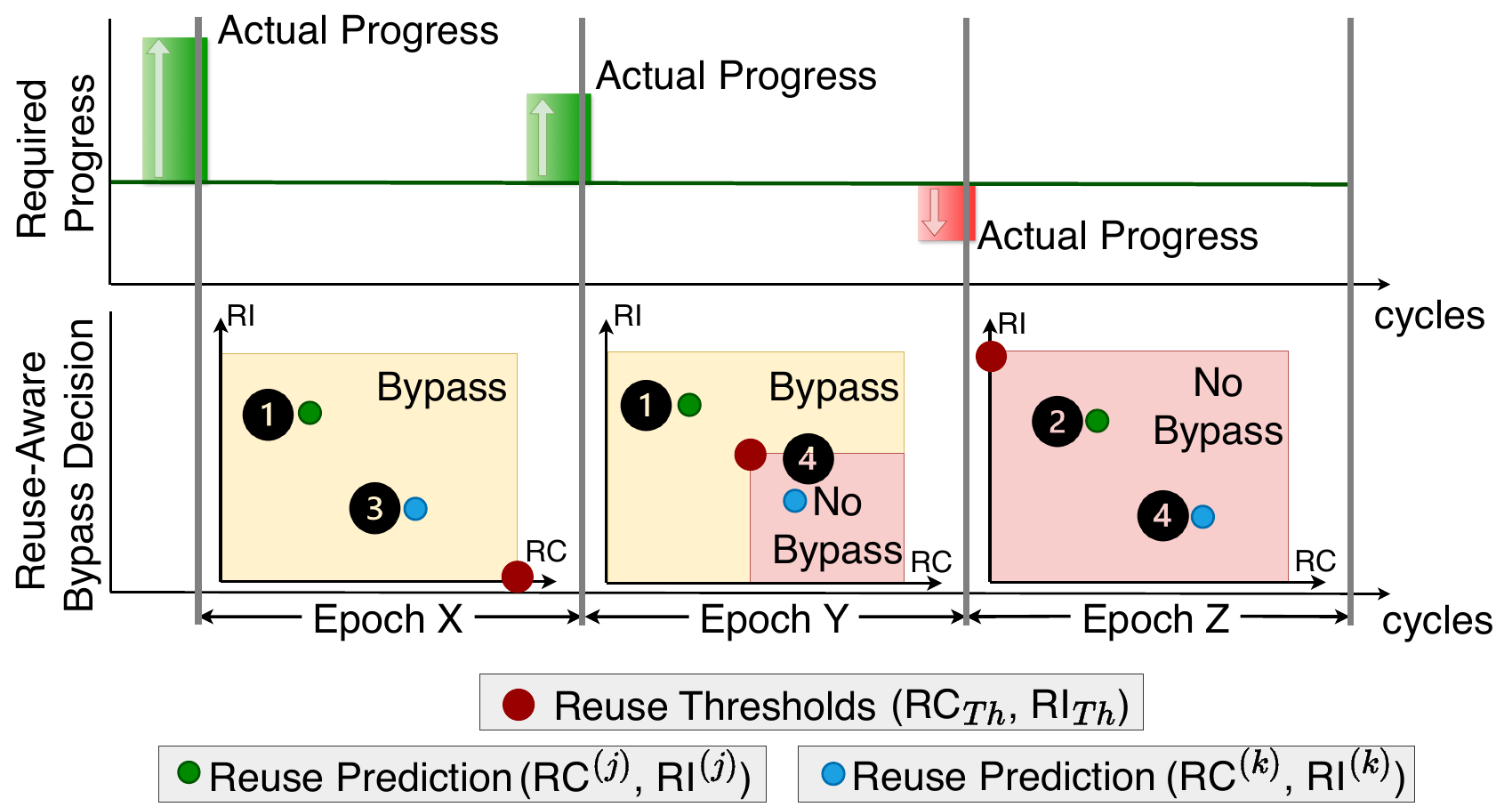}
        \caption{Overview of HyDRA's Bypass Conditions. Based on the actual progress of the accelerator and the reuse predictions, accesses are bypassed at the LLC.}
        \label{fig:hydra-bypass}
    \end{subfigure}
    \begin{subfigure}[b]{\linewidth}
        \centering
        \includegraphics[width=\linewidth]{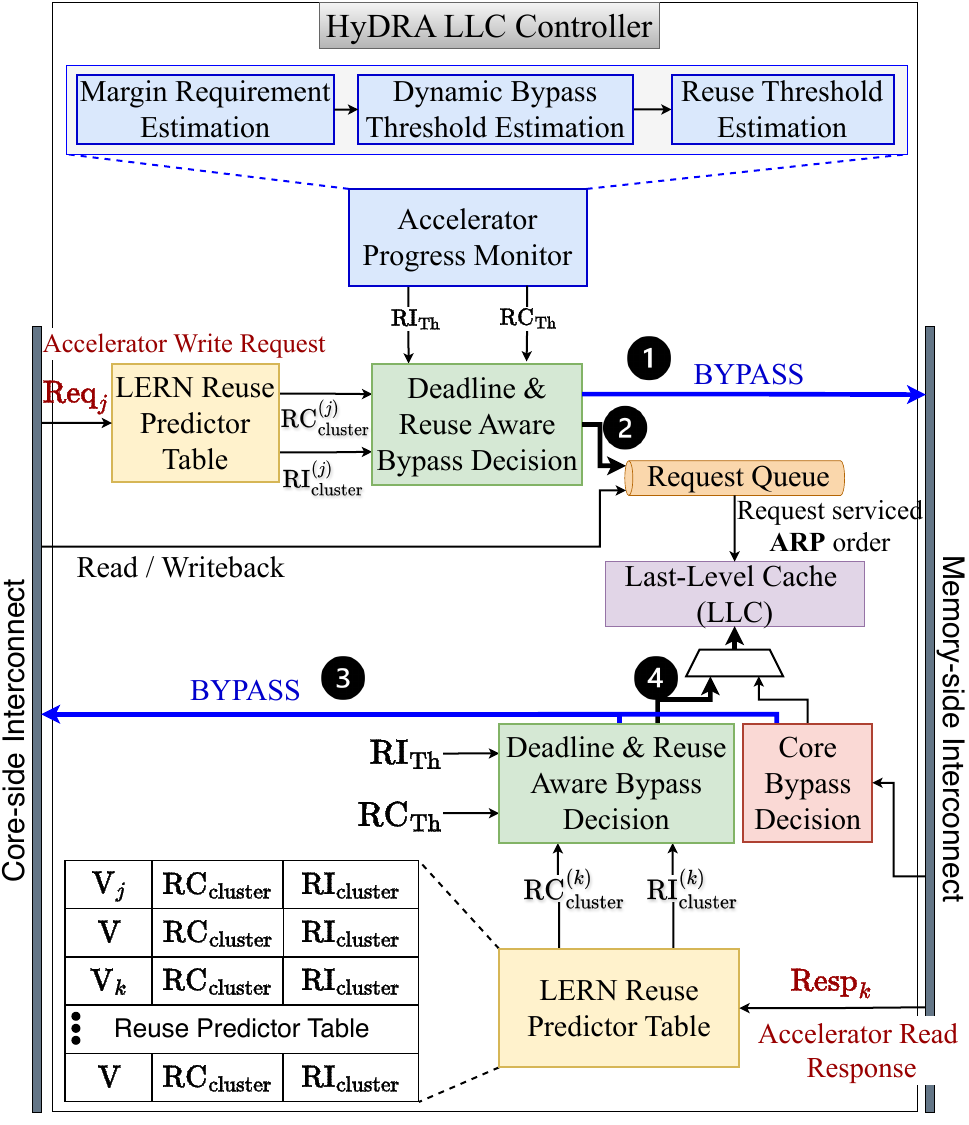}
        \caption{HyDRA LLC Controller Overview.}
        \label{fig:hydra-overview}
    \end{subfigure}
    \caption{HyDRA's Deadline with Reuse-Aware Bypass and Methodology Overview.}
\end{figure}

As motivated in Section \ref{sec:motivation}, with an accelerator-friendly arbitration policy, such as ARP, adding deadline awareness to reuse-aware cache bypass can not only help in meeting the accelerator's deadline but can also improve the system's throughput. 
Figure \ref{fig:hydra-overview} presents an overview of our proposed deadline and reuse-aware cacheability policy, HyDRA, which combines deadline awareness with \emph{dynamic reuse-aware cache bypass} and \emph{static bandwidth allocation by ARP arbitration} to achieve the twofold performance objective. 
It uses the LERN reuse predictor designed for the accelerator's memory access pattern to identify the bypass contenders based on reuse. 

As discussed in Section \ref{sec:deadlineAwareness}, the execution of the applications is divided into equal-sized epochs of length ET so that the state of the system and the progress of the accelerator are periodically monitored. 
The LLC management is adapted to the dynamic variations in the memory access patterns of the applications and the accelerator. 
Figure \ref{fig:hydra-bypass} illustrates the main ideas behind the reuse and deadline-aware bypass in HyDRA. The first plot shows whether the accelerator is ahead or behind the required progress, depending on its progress in the last epoch. The second plot shows the variation in the bypass decision and the access path taken at the LLC, according to the accelerator's progress and the predicted cache reuse of two accelerator accesses, $\mbox{Req}_j$ and $\mbox{Resp}_k$, as shown in Figure \ref{fig:hydra-overview}. When progress is good, we prefer to aggressively bypass accelerator access while taking a conservative bypass approach when the accelerator is behind the required progress. 
During epoch X, as the progress is excellent, both accesses are bypassed via access paths \tikz[baseline=(X.base)] \node (X) [draw, circle, inner sep=1pt, fill=black, text=white] {\small{1}}; and \tikz[baseline=(X.base)] \node (X) [draw, circle, inner sep=1pt, fill=black, text=white] {\small{3}}; as they lie in the bypass zone according to their predicted reuse. During epoch Y, actual progress is closer to the required progress, changing the reuse thresholds, $\mbox{RI}_{\mbox{\scriptsize Th}}$ and $\mbox{RC}_{\mbox{\scriptsize Th}}$. In this epoch, only $\mbox{Req}_j$ is bypassed and $\mbox{Resp}_k$ is sent to the LLC via access path \tikz[baseline=(X.base)] \node (X) [draw, circle, inner sep=1pt, fill=black, text=white] {\small{4}};.
However, during epoch Z, the progress is critical and hence, both accesses are sent to the LLC via access paths \tikz[baseline=(X.base)] \node (X) [draw, circle, inner sep=1pt, fill=black, text=white] {\small{2}}; and \tikz[baseline=(X.base)] \node (X) [draw, circle, inner sep=1pt, fill=black, text=white] {\small{4}};. Hence, with the dynamic exploitation of the tradeoff between deadline and reuse awareness, HyDRA can meet the deadline and enhance the cores' performance. 

HyDRA employs (1) an \textbf{Accelerator Progress Monitor (APM)} that monitors the progress of the accelerator at the beginning of every epoch to decide the reuse thresholds for bypassing memory accesses in this epoch, (2) a \textbf{pre-trained LERN Reuse Predictor Table} to dynamically predict the reuse behavior of the accelerator memory accesses, (3) a \textbf{Deadline- and Reuse-aware Bypass Decision} module which uses the prediction from LERN and the reuse thresholds to decide which memory accesses from the accelerator are bypassed at the LLC, and (4) a \textbf{Core Bypass Decision} module to bypass the read responses for the memory accesses from processor cores using a state-of-the-art reuse predictor. 

\subsection{Accelerator Progress Monitor (APM)}
\label{sec:apm}
As discussed in Section \ref{sec:deadlineAwareness}, the accelerator communicates a user-defined execution deadline (${D}_{\mbox{\scriptsize sec}}$) to the shared cache at the beginning of the execution along with the total memory accesses ${M}$ required to complete one input set on the accelerator. 
The Accelerator Progress Monitor (APM) uses these to estimate $\mbox{MA}_{\mbox{\scriptsize global}} = ({M}/{D}_{\mbox{\scriptsize sec}}) \times \mbox{ET} $, the total number of memory accesses which should ideally be completed every epoch to meet the accelerator's deadline. Based on the dynamic progress of the accelerator, the APM estimates the requirement of a safety margin to meet the deadline, the dynamic bypass thresholds, and finally, the reuse thresholds to be used for selecting the bypass candidates. 
At the beginning of each epoch $i$, APM checkpoints several performance counters for the accelerator, listed below, to estimate its progress and decide the reuse thresholds for the epoch. 
\begin{itemize}
    \item $\mbox{RA}^{(i)}$: Remaining accesses in this input set.
    \item $\mbox{RT}^{(i)}$: Remaining time to the deadline ${D}_{\mbox{\scriptsize sec}}$.
    \item $\mbox{MR}_i$: Total shared cache miss rate encountered by all cores in the last epoch $i-1$.
    \item $\mbox{MA}_{\mbox{\scriptsize past}}^{1\rightarrow i}$: The average number of memory accesses completed per epoch by the accelerator in the past until epoch $i-1$. It is defined as $(({M}-\mbox{RA}^{(i)}) \times \mbox{ET}) \div ({D}_{\mbox{\scriptsize sec}} - \mbox{RT}^{(i)}) $.
    \item $\mbox{AMAL}^{(i)}$: The average memory latency per access achieved by the accelerator in the last epoch $i-1$.
\end{itemize}

\subsubsection{Margin Requirement Estimation.}
\label{sec:margin-est}

\begin{figure}[htbp]
    \centering
    \includegraphics[width=\linewidth]{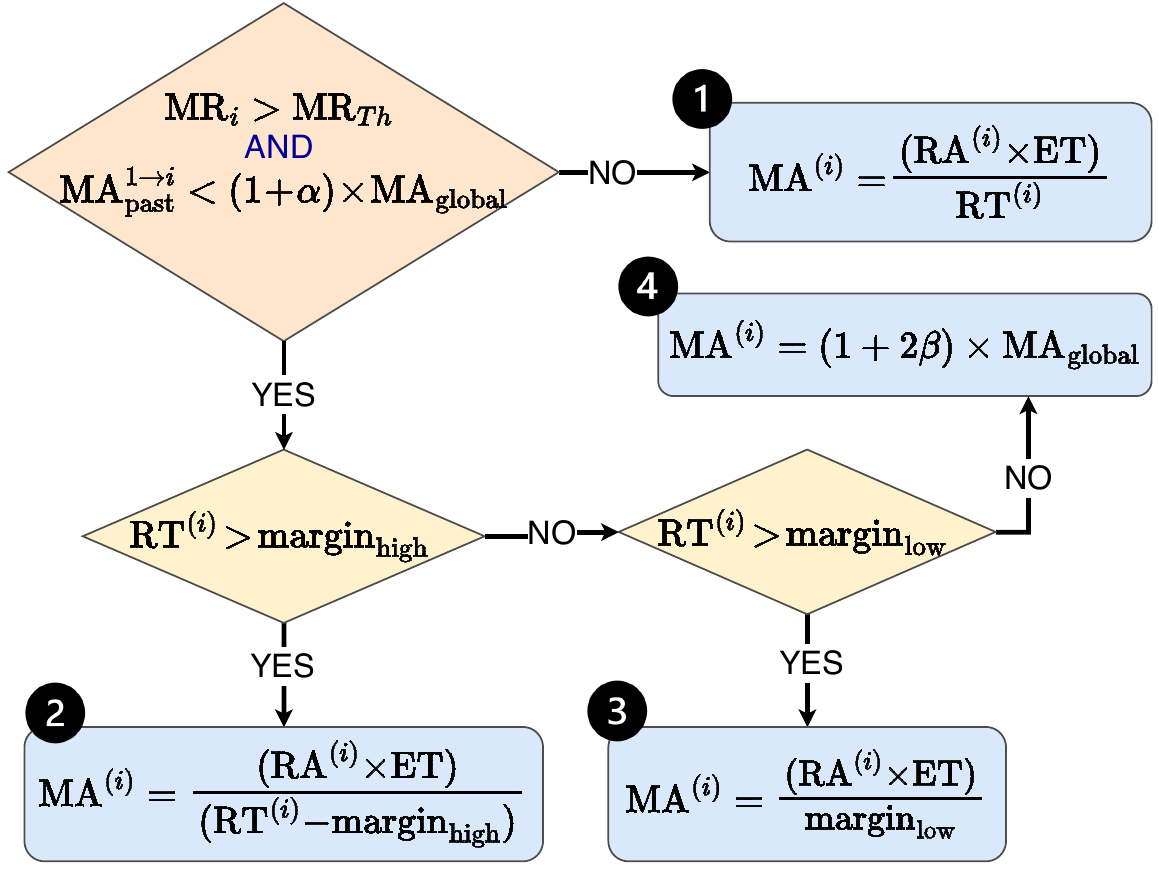}
     \caption{Margin Requirement Estimation based on the accelerator's progress and lower-level memory contention.}
    \label{fig:add-margin}
\end{figure}

Based on reuse behavior and deadline awareness, some accelerator accesses are selected to be bypassed from the shared cache to alleviate space contention and improve the system throughput. 
Aggressive bypass at the beginning of the accelerator execution can increase the contention at the lower-level memory or DRAM, leading to high per-access memory latency. If the accelerator cannot recover from the impact of the increased memory contention, it can lead to deadline misses. 
Therefore, the APM adds two safety margins, $\mbox{margin}_{\mbox{\scriptsize high}}$ and $\mbox{margin}_{\mbox{\scriptsize low}}$, in the deadline provided for the accelerator to recover from the impact of aggressive bypassing in cases when it is hard for the accelerator to meet the deadline. 
Figure \ref{fig:add-margin} shows that the safety margin requirement is estimated based on two conditions:
\begin{enumerate}[a)]
    \item If $\mbox{MR}_i > \mbox{MR}_{\mbox{\scriptsize Th}}$ (miss rate threshold), it shows higher existing contention at the DRAM. Aggressively bypassing the accelerator accesses can increase this contention, making it harder to meet the deadline. 
    \item If $\mbox{MA}^{1 \rightarrow i}_{\mbox{\scriptsize past}} < (1+\alpha) \times \mbox{MA}_{\mbox{\scriptsize global}}$, the APM foresees the requirement of a margin ($\alpha$ is empirical tolerance on global progress). This is because, with the current bypass strategy, the accelerator has not been able to meet the memory access completion rate required for every epoch. 
\end{enumerate}

Based on the margin requirement conditions shown in Figure \ref{fig:add-margin}, the per-epoch progress requirement or the number of accesses which are required to be completed in the epoch $i$, $\mbox{MA}^{(i)}$, is either estimated according to the actual remaining time from the deadline $\mbox{RT}_i$, as shown by condition \tikz[baseline=(X.base)] \node (X) [draw, circle, inner sep=1pt, fill=black, text=white] {\small{1}};, or gradually overestimated with addition of the margins to reduce the bypass aggressiveness, as shown by conditions \tikz[baseline=(X.base)] \node (X) [draw, circle, inner sep=1pt, fill=black, text=white] {\small{2}}; -- \tikz[baseline=(X.base)] \node (X) [draw, circle, inner sep=1pt, fill=black, text=white] {\small{4}};. 
We discuss in Section \ref{sec:ReuseThEst} how an inflated estimate of $\mbox{MA}^{(i)}$ reduces the bypass aggressiveness. 
{Even though the total work to be done for one input set is known when the accelerator begins execution, the number of accesses generated by the accelerator in every epoch during the execution is not known in advance. The dynamic, non-uniform, and unpredictable access rate from the accelerator plays a major role in meeting the per-epoch progress requirement. Low access rates in some epochs during execution can cause the accelerator to miss this epoch-level progress requirement $\mbox{MA}^{(i)}$. APM ensures the global deadline is met by adjusting the bypass aggressiveness in future intervals to compensate for deficits, as discussed in Sections \ref{sec:DynBypThresh} and \ref{sec:ReuseThEst}.} 

\subsubsection{Dynamic Bypass Threshold Estimation.}
\label{sec:DynBypThresh}

APM employs five dynamic bypass thresholds $\mbox{T}_{\mbox{\scriptsize A4}}^{(i)}$, $\mbox{T}_{\mbox{\scriptsize A3}}^{(i)}$, $\mbox{T}_{\mbox{\scriptsize A2}}^{(i)}$, $\mbox{T}_{\mbox{\scriptsize A1}}^{(i)}$, $\mbox{T}_{\mbox{\scriptsize B}}^{(i)}$ to determine the relative progress of the accelerator compared to the required epoch-level progress requirement $\mbox{MA}^{(i)}$ with $\beta$ tolerance, and decide the bypass aggressiveness for the accelerator's memory accesses in the form of dynamic reuse thresholds (Section \ref{sec:ReuseThEst}).
As shown in Algorithm \ref{alg:algo1}, at the beginning of each epoch $i$, after the margin requirement estimation, APM uses the proportional difference between $\mbox{MA}^{(i)}$ and $\mbox{MA}_{\mbox{\scriptsize global}}$, to increase or decrease the thresholds in multiples of step sizes $\delta_A$ and $\delta_B$, respectively. 
Higher values of the thresholds indicate less aggressive bypass, with reuse thresholds set such that the contenders for the bypass are limited, and lower values of these thresholds indicate a more aggressive bypass with reuse thresholds set to select bypass contenders aggressively.

\begin{algorithm}[t!]
    \caption{\textbf{Dynamic Bypass Threshold Estimation}: Update $\mbox{T}_{\mbox{\scriptsize A4}}^{(i)}$, $\mbox{T}_{\mbox{\scriptsize A3}}^{(i)}$, $\mbox{T}_{\mbox{\scriptsize A2}}^{(i)}$, $\mbox{T}_{\mbox{\scriptsize A1}}^{(i)}$, and $\mbox{T}_{\mbox{\scriptsize B}}^{(i)}$}
    \label{alg:algo1}
    \tcc{Dynamically update the bypass thresholds by monitoring the global progress of the accelerator on the current input set.}
    \tcp{Start of Epoch $i$}
    \Input{$\mbox{MA}^{(i)}$,  $\mbox{MA}_{\mbox{\scriptsize global}}$, $\beta, {\delta}_{\mbox{\scriptsize A}},  {\delta}_{\mbox{\scriptsize B}}$ }
    \Output{$\mbox{T}_{\mbox{\scriptsize A4}}^{(i)}$, $\mbox{T}_{\mbox{\scriptsize A3}}^{(i)}$, $\mbox{T}_{\mbox{\scriptsize A2}}^{(i)}$, $\mbox{T}_{\mbox{\scriptsize A1}}^{(i)}$, and $\mbox{T}_{\mbox{\scriptsize B}}^{(i)}$}
    \ForAll{$j \in \{1, 2, 3, 4\}$} {
    \If{$\mbox{\emph{MA}}^{(i)}\le (1\!-\!6\beta) \mbox{\emph{ MA}}_{\mbox{\scriptsize \emph{global}}}$}
        { $\mbox{T}_{\mbox{\scriptsize Aj}}^{(i)} \gets \max(\mbox{T}_{\mbox{\scriptsize Aj}}-6{\delta}_{\mbox{\scriptsize A}},1);\;
        \mbox{T}_{\mbox{\scriptsize B}}^{(i)} \gets \mbox{T}_{\mbox{\scriptsize B}}-6{\delta}_{\mbox{\scriptsize B}}$ 
        }
    \For{$k \gets 5$ \KwTo $1$}{
        \If{$
        (1-(k+1)\beta) \mbox{\emph{ MA}}_{\mbox{\scriptsize \emph{global}}} < \mbox{\emph{MA}}^{(i)} \;\mathbf{ and }\;$ 
        $\mbox{\emph{MA}}^{(i)} \le (1-k\beta) \mbox{\emph{ MA}}_{\mbox{\scriptsize \emph{global}}}
        $}
           { $\mbox{T}_{\mbox{\scriptsize Aj}}^{(i)} \leftarrow \max(\mbox{T}_{\mbox{\scriptsize Aj}} - k{\delta}_{\mbox{\scriptsize A}},1)$
           
           $\mbox{T}_{\mbox{\scriptsize B}}^{(i)} \leftarrow \mbox{T}_{\mbox{\scriptsize B}} - k{\delta}_{\mbox{\scriptsize B}}$ }
    }
    \If{$
    (1\!-\!\beta) \mbox{\emph{ MA}}_{\mbox{\scriptsize \emph{global}}} < \mbox{\emph{MA}}^{(i)} \le (1\!+\!\beta) \mbox{\emph{ MA}}_{\mbox{\scriptsize \emph{global}}}
    $}
       { $\mbox{T}_{\mbox{\scriptsize Aj}}^{(i)} \leftarrow \mbox{T}_{\mbox{\scriptsize Aj}};\;
       \mbox{T}_{\mbox{\scriptsize B}}^{(i)} \leftarrow \mbox{T}_{\mbox{\scriptsize B}}$ }
    \If{$ \mbox{\emph{MA}}^{(i)} > (1\!+\!\beta) \mbox{\emph{ MA}}_{\mbox{\scriptsize \emph{global}}}
    $}
       { $\mbox{T}_{\mbox{\scriptsize Aj}}^{(i)} \leftarrow \mbox{T}_{\mbox{\scriptsize Aj}} + {\delta}_{\mbox{\scriptsize A}};\;
       \mbox{T}_{\mbox{\scriptsize B}}^{(i)} \leftarrow \mbox{T}_{\mbox{\scriptsize B}}$ }
    }
\end{algorithm}

\subsubsection{Reuse Threshold Estimation.}
\label{sec:ReuseThEst}

For any epoch $i$, $\mbox{MA}^{(i)}$ serves as the accelerator's progress requirement. 
The APM also monitors the average per-access memory latency, $\mbox{AMAL}^{(i)}$, achieved by the accelerator in the last epoch $i-1$, which denotes the time each accelerator access spends in the memory hierarchy.
APM uses this latency to predict the number of memory accesses that the accelerator would be able to complete in this epoch $i$ or $\widehat{\mbox{MA}}^{(i)} = \mbox{ET}/\mbox{AMAL}^{(i)}$, given the same shared cache management strategy as the last epoch.

APM uses the five dynamic bypass thresholds, discussed in Section \ref{sec:DynBypThresh}, to determine the difference proportion between $\widehat{\mbox{MA}}^{(i)}$ and $\mbox{MA}^{(i)}$ and estimate how far ahead or behind the accelerator is in this epoch $i$. 
If $\widehat{\mbox{MA}}^{(i)}\! > \! \mbox{MA}^{(i)}$, it is likely that the accelerator would complete the required accesses, with a possibility of increasing its bypass aggressiveness. 
If $\widehat{\mbox{MA}}^{(i)} \! < \! \mbox{MA}^{(i)}$, accelerator-friendly changes like reducing the bypass aggressiveness have to be made to meet the progress requirement. 
The bypass aggressiveness is controlled by reuse thresholds, which control access-level bypass according to the reuse behavior, instead of regulating the percentage of accesses bypassed at the LLC controller. 
If progress permits, accesses with good reuse properties can also be bypassed to improve cores' performance. 
This contradicts the common reuse-aware notion without deadline awareness that good reuse accesses are prime candidates for caching. 

\begin{figure}[b!]
    \centering
    \includegraphics[width=\linewidth]{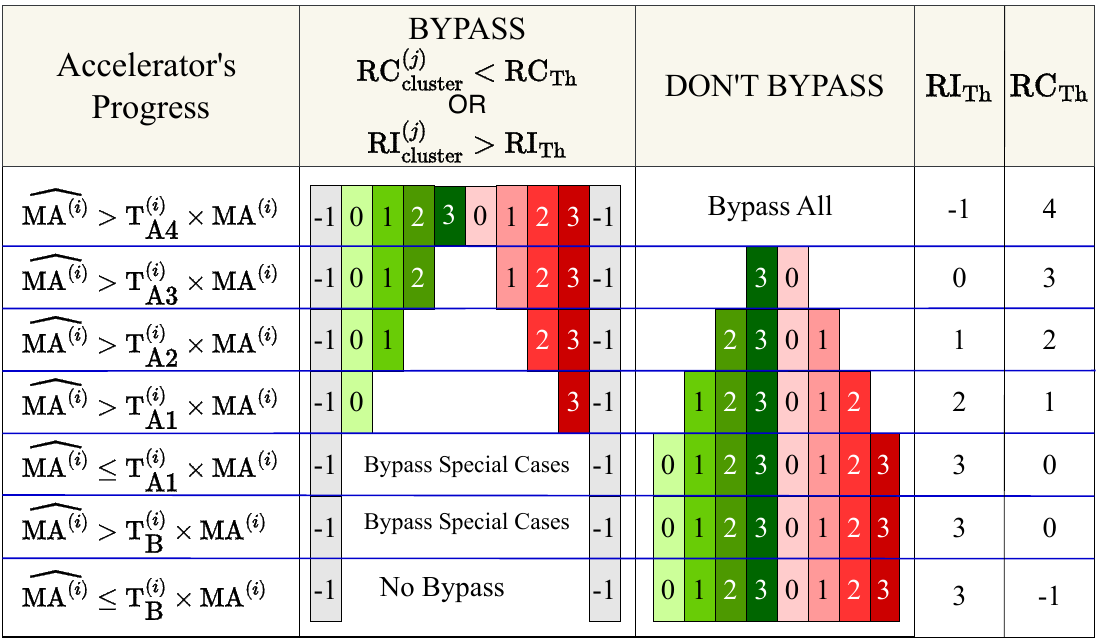}
    \caption{Clusters to be bypassed determined by the reuse thresholds $\mbox{RC}_{\mbox{\scriptsize Th}}$ and $\mbox{RI}_{\mbox{\scriptsize Th}}$ selected based on the progress.}
    \label{fig:reuse-threshold}
\end{figure}

Figure \ref{fig:reuse-threshold} shows the selection of the reuse thresholds, $\mbox{RI}_{\mbox{\scriptsize Th}}$ and $\mbox{RC}_{\mbox{\scriptsize Th}}$, based on the progress estimate, discussed above. At the LLC controller, the bypass decision module dynamically bypasses all accesses in cluster $\mbox{RI}_{\mbox{\scriptsize cluster}}$, $\mbox{RC}_{\mbox{\scriptsize cluster}}$ (LERN-clusters) if $\mbox{RI}_{\mbox{\scriptsize cluster}} > \mbox{RI}_{\mbox{\scriptsize Th}}$ or $\mbox{RC}_{\mbox{\scriptsize cluster}} < \mbox{RC}_{\mbox{\scriptsize Th}}$ (Section \ref{sec:DRABypDecision}). 
According to the proportional difference between $\widehat{\mbox{MA}}^{(i)}$ and $\mbox{MA}^{(i)}$ being within certain bypass threshold levels, the clusters whose accesses will be bypassed at the LLC are decided by varying the reuse thresholds between -1 to 4 to increase or decrease the bypass aggressiveness. If sufficient progress has been made, more cache space is reallocated to the cores by choosing reuse thresholds to bypass accelerator accesses aggressively. 
Depending on the progress, the reuse thresholds are modified to vary the bypass aggressiveness. 
Beyond $\mbox{T}_{\mbox{\scriptsize  A1}}^{(i)}$ and $\mbox{T}_{\mbox{\scriptsize  B}}^{(i)}$, HyDRA chooses the reuse thresholds such that no accesses are bypassed, as the deadline could be missed, apart from some special cases discussed in Section \ref{sec:DRABypDecision}.

\subsection{LERN Reuse Predictor Table}
\label{sec:lernPredict}

For each machine learning model and accelerator configuration, offline learning of reuse patterns is performed to obtain a memory access-to-cluster mapping using the LERN methodology (Section \ref{sec:lern}). 
{We propose a hybrid technique in which complex reuse behavior is learnt offline while compact hardware tables perform lightweight cluster-address mapping at runtime.
At the LLC controller, we use a tagless, direct-mapped buffer called the LERN reuse predictor table (L-RPT) to store this mapping for the machine learning model, layer by layer. It has 512K entries, with 5 bits per entry (2 bits each for RI and RC-cluster IDs, and 1 valid bit), indexed by the lower block address bits (memory address $>>$ 6). 
The cluster information for all block addresses in a single layer that show reuse during offline training is populated in this table. An invalid table entry represents an access with ``No Reuse". 
Each valid L-RPT lookup returns the cluster allocations for that block address, learnt offline. 
The hardware optimizations and overheads of the predictor table are discussed in Sections \ref{sec:lrptOpt} and \ref{sec:overhead}. 
}

\subsection{Deadline and Reuse-Aware Bypass Decision}
\label{sec:DRABypDecision}

The deadline and reuse-aware bypass decision module uses $\mbox{RI}_{\mbox{\scriptsize Th}}$ and $\mbox{RC}_{\mbox{\scriptsize Th}}$ along with the dynamic per-access reuse predictions from the LERN predictor or L-RPT to decide the bypass contenders at LLC. 
As shown in Figure \ref{fig:hydra-overview}, when a write request $\mbox{Req}_j$ from the accelerator arrives at the LLC, a lookup is performed in the L-RPT to find a valid entry for this request. The L-RPT returns a reuse prediction or cluster allocation for this request in the form of a tuple, $(\mbox{RC}^{(j)}_{\mbox{\scriptsize cluster}}, \mbox{RI}^{(j)}_{\mbox{\scriptsize cluster}})$.
The request is bypassed if $\mbox{RI}^{(j)}_{\mbox{\scriptsize cluster}} > \mbox{RI}_{\mbox{\scriptsize Th}}$ or $\mbox{RC}^{(j)}_{\mbox{\scriptsize cluster}} < \mbox{RC}_{\mbox{\scriptsize Th}}$. {If the cluster allocation is to the No Reuse cluster (-1) or an invalid table entry, $\mbox{Req}_j$ is bypassed from the LLC, suggesting no further reuse.} If the request does not meet the bypass criteria, it is added to the LLC request queue and is serviced at priority by the LLC (ARP static bandwidth partitioning). 
Figure \ref{fig:reuse-threshold} provides a summary of the bypass decisions based on the reuse predictions and the reuse thresholds. 
If the write request $\mbox{Req}_j$ chosen for bypass is already present in the LLC, the cache copy is invalidated before proceeding to the next request. 
For a read request from the accelerator $\mbox{Req}_k$ already in the LLC, the bypass decision is ignored, and the cache services the request. In case of a miss, the L-RPT is looked up when the response $\mbox{Resp}_k$ arrives at the cache from the DRAM. The bypass decision criteria are similar to those for a write request bypass.    
As shown in Figure \ref{fig:reuse-threshold}, when $\mbox{RC}_{\mbox{\scriptsize Th}} = 0$ and the progress of the accelerator is critical, HyDRA conservatively bypasses accesses (special cases) in the Cold RC-cluster only if the cluster center is such that the accesses might only be reused only once more. All the other accesses are not bypassed.

\subsection{Core Bypass Decisions}
\label{sec:coreByp}
Previous works have shown that reuse-aware bypassing of memory accesses from cores at the shared cache can alleviate space and bandwidth contention. We use state-of-the-art reuse predictors such as SHIP~\cite{SHIP} to select the bypass contenders from processor cores. The choice of the predictor is not limited to SHIP, and any other predictor can replace it. A reuse predictor table learns the reuse behavior of memory accesses based on the past and selects the accesses with poor reuse behavior for bypass. 
This further alleviates the LLC space and bandwidth contention and creates a positive feedback loop between the core and accelerator bypass. Core bypass leads to better accelerator progress, which increases its progress-dependent bypass aggressiveness and improves the cores' performance. 

The parameter selection and the overheads of HyDRA are discussed in Section \ref{sec:paramSel} and Section \ref{sec:overhead}, respectively.

\section{Experiments and Results}
\label{sec:experiments}

\subsection{Infrastructure Enhancements to gem5}
\label{sec:Infra}

Cycle-accurate memory access traces extracted from accelerator simulators, such as SCALE-Sim~\cite{samajdar2020systematic}, and integrated into the gem5 architectural simulator, reasonably model the shared cache contention between hardware accelerators and processor cores~\cite{FLASH}. 
By default, all the memory accesses arriving at the shared cache are written into the cache (\emph{black} paths in Figure \ref{fig:introduction}). 
We enhance this infrastructure to add selective bypassing of the accelerator's memory accesses from the shared cache. 
We add a direct BYPASS request path for write requests from the requester to the DRAM, and a BYPASS response path for both read and write responses from the DRAM to the requester (highlighted in \emph{blue} in Figure \ref{fig:introduction}). 

\begin{table}[htbp]
  \caption{System Configuration.}
  \label{tab:systemconfig}
  \begin{tabularx}{\linewidth}{>{\raggedright\arraybackslash}p{1.95cm}|>{\raggedright\arraybackslash}p{6.02cm}}
    \hline
    \rowcolor{gray!20} \textbf{Component} & \multirow{1}{*}{\textbf{Configuration}}\\
    \hline
    \hline
    \multirow{3}{*}{CPU }& 8 Intel X86 Out-of-Order (OoO) cores, 2 GHz,{ROB size: 192, LQ/SQ entries: 32/32, Fetch/Decode/Issue/Commit/WB Width: 8/8/8/8/8}\\
    \hline
    \rowcolor{gray!20} \multirow{2}{*}{L1 Cache} & Private, 32KB L1-D, 32KB L1-I, 64B block, 8-way, parallel-access, Tag: 1 cycle, Data: 3 cycles \\
    \hline
    \multirow{2}{*}{L2 Cache} & Private, 256KB, 64B block, 8-way, parallel-access, Tag: 2 cycles, Data: 5 cycles \\
    \hline
    \rowcolor{gray!20} \multirow{3}{*}{Last-Level Cache} & Shared, {Non-sliced}, 8MB, 64B block, 16-way, parallel-access, Non-Inclusive, Tag: 3 cycles, Data: 9 cycles \\
    \hline
  \multirow{2}{*}{DRAM} & 8GB DDR3, 1600MHz, single channel, 2 ranks/channel, 8 banks/rank, page size: 1KB\\
  \hline
  \rowcolor{gray!20} \multirow{2}{*}{Accelerator} & Systolic PE Array, double-buffered On-Chip SRAM, OS/WS/IS dataflow (DF)\\
    \hline
\end{tabularx}
\end{table}

\begin{table}[htbp]
  \caption{SPEC CPU2006 benchmark evaluation workloads. Abbreviation: \textcolor{black}{gamess (gs), soplex (so), omnetpp (om).}} 
  \label{tab:specmixes}
  \begin{tabularx}{\linewidth}{>{\raggedright\arraybackslash}p{0.7cm}|>{\raggedright\arraybackslash}p{7.27cm}}
    \hline
    \rowcolor{gray!20} \textbf{Mixes} & \multicolumn{1}{c}{\textbf{Benchmarks running on 8 cores}}\\
    \hline
    \hline
   \multirow{1}{*}{Mix 1} & \textcolor{teal}{{wrf}\textcolor{black}{,} {hmmer}\textcolor{black}{,} {gromacs}\textcolor{black}{,} {namd}\textcolor{black}{,} {bzip2}\textcolor{black}{,} {gromacs}\textcolor{black}{,} {povray}\textcolor{black}{,} {dealII}} \\
    \hline
   \rowcolor{gray!20} \multirow{1}{*}{Mix 2} & \textcolor{violet}{{so}\textcolor{black}{,} {so}\textcolor{black}{,} so\textcolor{black}{,} so\textcolor{black}{,} gs\textcolor{black}{,} gs\textcolor{black}{,} om\textcolor{black}{,} om} \\
    \hline
   \multirow{1}{*}{Mix 3} & \textcolor{violet}{{gs}\textcolor{black}{,} {gs}\textcolor{black}{,} {gs}\textcolor{black}{,} so\textcolor{black}{,} so\textcolor{black}{,} om\textcolor{black}{,} om\textcolor{black}{,} om}\\
    \hline
   \rowcolor{gray!20} \multirow{1}{*}{Mix 4} &  \textcolor{violet}{{so}\textcolor{black}{,} {gs}\textcolor{black}{,} so\textcolor{black}{,} om\textcolor{black}{,} so\textcolor{black}{,} gs\textcolor{black}{,} gs\textcolor{black}{,} gs} \\
    \hline
   \multirow{1}{*}{Mix 5} & \textcolor{violet}{{om}\textcolor{black}{,} {om}\textcolor{black}{,} {so}\textcolor{black}{,} gs\textcolor{black}{,} gs\textcolor{black}{,} gs\textcolor{black}{,} so\textcolor{black}{,} so} \\
    \hline
   \rowcolor{gray!20} \multirow{1}{*}{Mix 6}  & \textcolor{teal}{Gems{\tiny{FDTD}}\textcolor{black}{,}	hmmer\textcolor{black}{,}	Gems{\tiny{FDTD}}\textcolor{black}{,}} \textcolor{violet}{	gs\textcolor{black}{,}} \textcolor{purple}{	bwaves\textcolor{black}{,}	lbm\textcolor{black}{,}	mcf\textcolor{black}{,} zeusmp} \\
    \hline
  \multirow{1}{*}{Mix 7} &  \textcolor{teal}{povray\textcolor{black}{,}	astar\textcolor{black}{,}	gromacs\textcolor{black}{,}} \textcolor{violet}{	om\textcolor{black}{,}	gs\textcolor{black}{,}	om\textcolor{black}{,}	so\textcolor{black}{,}	gs} \\
    \hline
   \rowcolor{gray!20} Mix 8 &  \textcolor{teal}{sjeng\textcolor{black}{,} namd\textcolor{black}{,}	gobmk\textcolor{black}{,}	bzip2\textcolor{black}{,}} \textcolor{purple}{	lbm\textcolor{black}{,}	bwaves\textcolor{black}{,} libquantum\textcolor{black}{,}	mcf} \\
    \hline
    \multirow{1}{*}{Mix 9}  & \textcolor{violet}{gs\textcolor{black}{,}	gs\textcolor{black}{,}	gs\textcolor{black}{,}	so\textcolor{black}{,}	om\textcolor{black}{,} }\textcolor{purple}{	mcf\textcolor{black}{,}	milc\textcolor{black}{,}	zeusmp}\\
    \hline
    \rowcolor{gray!20} \multirow{1}{*}{Mix 10}&  \textcolor{teal}{povray\textcolor{black}{,}	dealII\textcolor{black}{,}} \textcolor{violet}{	so\textcolor{black}{,} om\textcolor{black}{,} gs\textcolor{black}{,}	gs\textcolor{black}{,}} \textcolor{purple}{lbm\textcolor{black}{,}	milc}\\
    \hline
    \multirow{1}{*}{Mix 11} &  \textcolor{teal}{hmmer\textcolor{black}{,}	hmmer\textcolor{black}{,}} \textcolor{violet}{gs\textcolor{black}{,}	gs\textcolor{black}{,}} \textcolor{purple}{	lbm\textcolor{black}{,} milc\textcolor{black}{,}	leslie3d\textcolor{black}{,}	bwaves} \\
    \hline
    \rowcolor{gray!20} \multirow{1}{*}{Mix 12} & \textcolor{teal}{h264ref\textcolor{black}{,}} \textcolor{violet}{	gs\textcolor{black}{,} so\textcolor{black}{,} gs\textcolor{black}{,}	so\textcolor{black}{,}} \textcolor{purple}{ mcf\textcolor{black}{,} lbm\textcolor{black}{,}	zeusmp}\\
    \hline
  \end{tabularx}
\end{table}

\begin{table}[htbp]
  \caption{Hardware Accelerator Configurations. }
  \centering
  \label{tab:hwaconfig}
  \begin{tabularx}{\linewidth}{>{\centering\arraybackslash}p{1cm}|>{\centering\arraybackslash}p{1.5cm}|>{\centering\arraybackslash}p{1.13cm}|>{\centering\arraybackslash}p{2.73cm}|>{\centering\arraybackslash}p{0.3cm}}
    \hline
    \multirow{3}{*}{\textbf{Config-\emph{n}}} &  \multirow{3}{*}{\textbf{ML model}} & \multicolumn{3}{c}{\textbf{Accelerator Parameters}} \\
    \cline{3-5}
    & & \multirow{2}{*}{\textbf{PE Array}} & \multicolumn{1}{c|}{\textbf{SRAM size (in KB)}} & \multirow{2}{*}{\textbf{DF}}  \\
    \cline{4-4}
    & &  & IFMAP, OFMAP, FIL &  \\
    \hline
    \hline
    \rowcolor{gray!20} \multirow{1}{*}{Config-1} & \multirow{1}{*}{Tiny-YOLO} &  $256 \times 256$  & 6144, 6144, 6144 &  OS  \\
    \hline
    \multirow{1}{*}{Config-2} & \multirow{1}{*}{Tiny-YOLO} &  $256 \times 256$ & 6144, 6144, 6144 &  {WS } \\
    \hline
    \rowcolor{gray!20} \multirow{1}{*}{Config-3} & \multirow{1}{*}{{Tiny-YOLO}}  &  $256 \times 256$ & 64, 64, 64 & OS  \\
    \hline
    \multirow{1}{*}{Config-4} & \multirow{1}{*}{Tiny-YOLO} &  $64 \times 64$ & 64, 64, 64 & OS  \\
    \hline
    \rowcolor{gray!20} \multirow{1}{*}{Config-5} & \multirow{1}{*}{GoogleNet} & $64 \times 64$&  64, 64, 64 & OS  \\
    \hline
    \multirow{1}{*}{Config-6} & \multirow{1}{*}{GoogleNet} & $64 \times 64$&  64, 64, 64 & WS  \\
    \hline
    \rowcolor{gray!20} \multirow{1}{*}{Config-7} & \multirow{1}{*}{MobileNet} & $64 \times 64$ &  64, 64, 64 & OS  \\
    \hline
    \multirow{1}{*}{Config-8} & \multirow{1}{*}{Deep Speech2} & $64 \times 64$ &  64, 64, 64 & OS  \\
    \hline
    \rowcolor{gray!20} \multirow{1}{*}{Config-9} & \multirow{1}{*}{Faster R-CNN} & $256 \times 256$&  6144,   6144,  6144&  OS  \\
    \hline
    \multirow{1}{*}{Config-10} & \multirow{1}{*}{AlphaGoZero} & $64 \times 64$ &  64, 64, 64 & OS  \\
    \hline
  \end{tabularx}
\end{table}

\begin{figure*}[b!]
    \centering
    \begin{subfigure}[b]{\linewidth}
        \includegraphics[width=\linewidth]{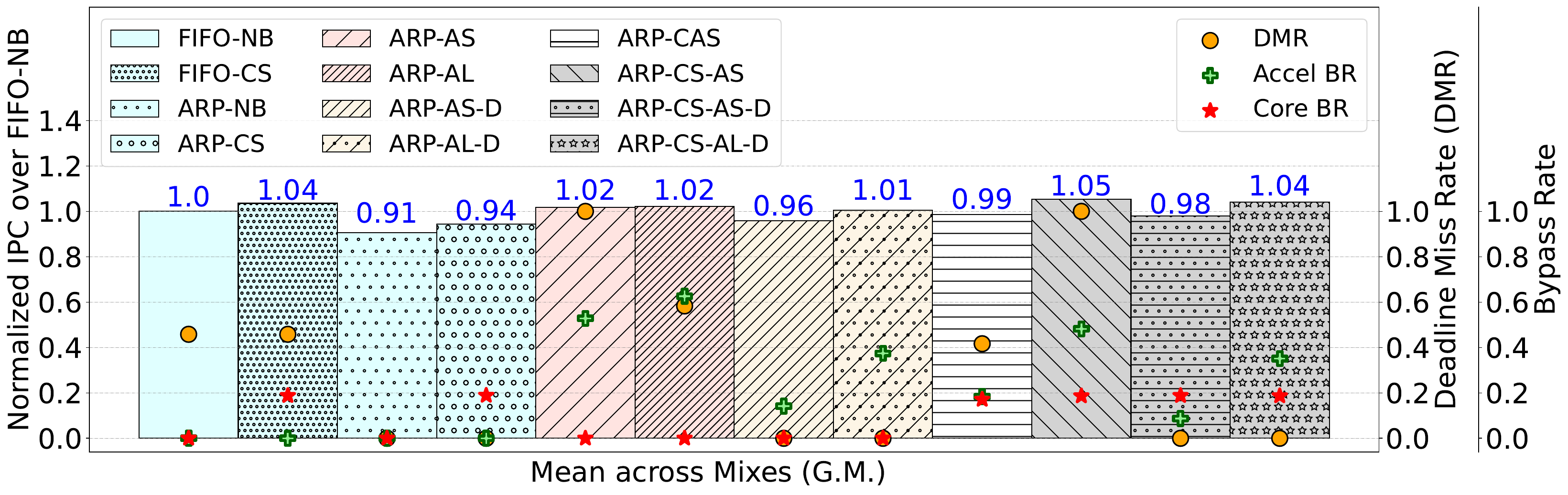}
        \caption{{Performance Comparison of HyDRA with different cache space management policies.}}
        \label{fig:config1-all}
    \end{subfigure}
    \par \vspace{1em}
    \begin{subfigure}[b]{\linewidth}
        \includegraphics[width=\linewidth]{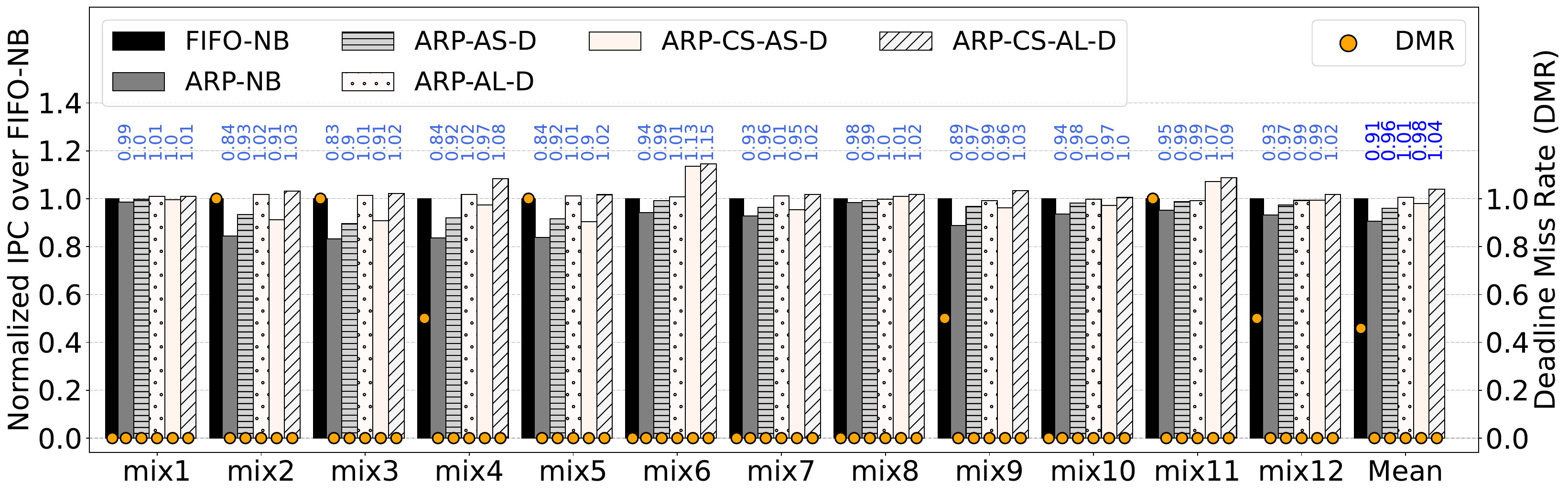}
        \caption{{Performance Evaluation of HyDRA across different workload mixes.}}
        \label{fig:config1-main}
    \end{subfigure}
    \caption{{Performance Evaluation of HyDRA (ARP-CS-AL-D) on Accelerator Config-1. Deadline: 10 IPS.}}
    \label{fig:config1}
\end{figure*}

\subsection{Experimental Setup}
\label{sec:setup}

{We experimentally evaluate the performance of HyDRA with the enhanced gem5 architectural simulator in system call emulation mode (SE mode) on a system (Table \ref{tab:systemconfig}) representative of modern heterogeneous edge systems, including networking, automotive, aerospace, mobile, IoT, and edge-AI SoCs.}
{We choose workloads from the SPEC CPU2006 benchmark suite and categorize them into \textcolor{teal}{{Compute- (CI)}}, \textcolor{violet}{{LLC- (LI)}}, or \textcolor{purple}{{Memory-Intensive (MI)}} based on their standalone cache behavior.}
We create two workload mixes using applications from different categories. 
A set of 30 workload mixes determines the policy parameters (Section \ref{sec:paramSel}). Once the parameters are finalized, a different set of 12 workload mixes, called the \emph{evaluation set} (Table \ref{tab:specmixes}), is used for the final evaluation.
For each workload, we fast-forward for 1 billion instructions with the accelerator idle, then run the final simulation with the accelerator, which terminates when any core completes 400 million instructions. 

We use SCALE-Sim~\cite{samajdar2020systematic} to generate the cycle-accurate memory access traces for different accelerator configurations (Table \ref{tab:hwaconfig}). {These configurations cover a broad design space with different machine learning models such as Tiny-YOLO, GoogleNet, MobileNet, Deep Speech2, Faster R-CNN, and AlphaGoZero, simulated on accelerator architectures with different SRAM, PE array sizes, and dataflow (DF) such as Output-Stationary (OS), Input-Stationary (IS), or Weight-Stationary (WS).}

\subsection{Performance Evaluation}
\label{sec:perfEval}

We measure the overall system throughput using the combined Instructions Per Cycle (IPC) achieved by all eight processor cores. 
We use the Deadline Miss Rate (DMR) metric, defined in Section \ref{sec:motivation}, to determine accelerators' performance.
We use a deadline derived from an input rate of 10 IPS (inputs: frames/words/tokens) in our evaluation. {The policy \textbf{ARP-CS-AL-D} is abbreviated as HyDRA, and \textbf{ARP-AL-D} represents a variation of HyDRA without core bypass.}

\subsubsection{Baselines.}
Since shared cache space management between accelerators and cores has not been studied in previous work, we compare HyDRA with baselines based on state-of-the-art reuse predictors and cache bypass. We use the SHIP~\cite{SHIP} as the baseline predictor for the cores and the accelerator. {Deadline-aware SHIP-driven accelerator bypass begins once the epoch-level progress requirement is met (Section \ref{sec:deadlineAwareness}). Bypass policies in the notations are denoted by their corresponding first letter (S: SHIP, L: LERN Predictor).}
\begin{enumerate}[a)]
    \item {FIFO/ARP-NB and FIFO/ARP-CS: The requests at the LLC are served in FIFO/ARP order with no bypass (NB) and with \textbf{S}HIP-driven reuse-aware \textbf{C}ore bypass (CS).}
    \item {ARP-AS/ARP-AS-D: The LLC requests are served in ARP order with deadline-agnostic or deadline-aware ($-D$) \textbf{S}HIP-driven \textbf{A}ccelerator bypass (AS).}
    \item {ARP-AL: The LLC requests are served in ARP order with deadline-agnostic reuse-aware \textbf{A}ccelerator bypass using the pre-trained \textbf{L}ERN predictor (AL).} 
    \item {ARP-CAS: The LLC requests are served in ARP order with a deadline-agnostic, reuse-aware \textbf{C}ore and \textbf{A}ccelerator bypass, using a shared \textbf{S}HIP predictor (CAS).}
    \item {ARP-CS-AS/ARP-CS-AS-D: The LLC requests are served in ARP order with deadline-agnostic (or deadline-aware) reuse-aware accelerator and core bypass with different SHIP predictors (CS-AS).}
\end{enumerate}

\begin{figure*}[b!]
    \centering
    \includegraphics[width=\linewidth]{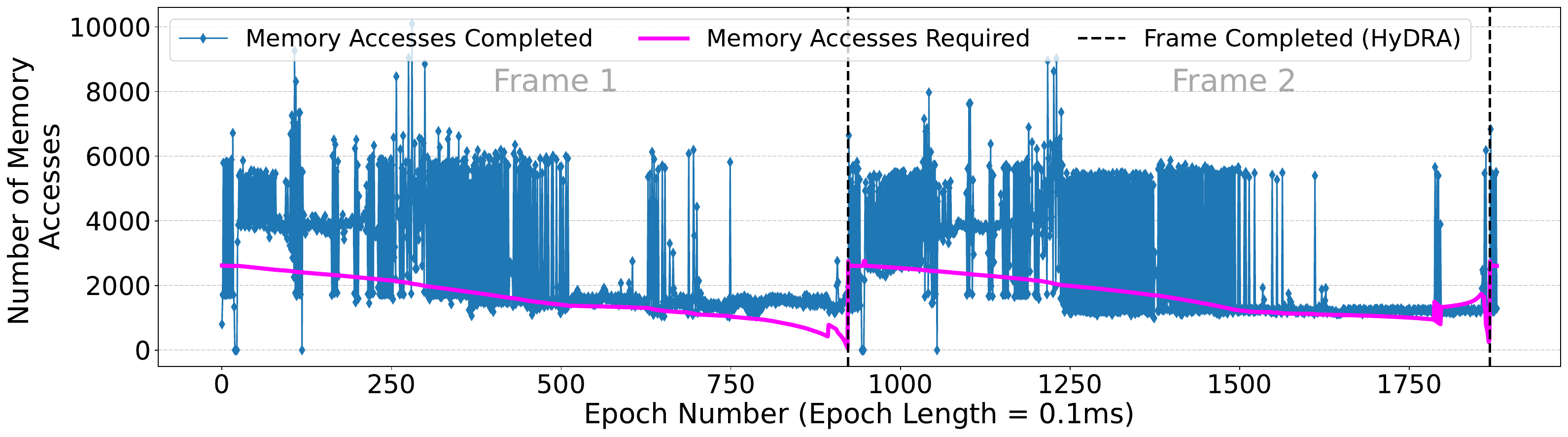}
    \caption{Accelerator's shared cache access rate during the execution compared with the per-epoch progress requirement based on deadline. Accelerator Configuration: Config-1, \emph{mix4}. Deadline: 10 IPS.}
    \label{fig:mix4AccessRate}
\end{figure*}

\subsubsection{Performance breakdown of HyDRA and all the baselines on Accelerator Config-1.}
\label{sec:config1}

Figure \ref{fig:config1-all} shows the performance of all the policies on Config-1 compared to the FIFO-NB baseline. {It shows the accelerator bypass (Accel BR) and core bypass (Core BR) rates, marked in green and red, respectively.}

As discussed in Section \ref{sec:keyInsight1}, we observe that \tikz[baseline=(X.base)] \protect\node (X) [draw, circle, fill=white, text=black, inner sep=1.5pt] {\small{1}}; FIFO-NB is unable to provide sufficient bandwidth to the accelerator to meet the deadline, \tikz[baseline=(X.base)] \protect\node (X) [draw, circle, fill=white, text=black, inner sep=1.5pt] {\small{2}}; FIFO-CS bypasses up to 20\% core accesses and gives 4\% IPC speedup, but does not alleviate enough contention to meet the deadline, \tikz[baseline=(X.base)] \protect\node (X) [draw, circle, fill=white, text=black, inner sep=1.5pt] {\small{3}}; ARP-NB meets the deadline but leads to a 9\% drop in IPC, \tikz[baseline=(X.base)] \protect\node (X) [draw, circle, fill=white, text=black, inner sep=1.5pt] {\small{4}}; ARP-CS bypasses 20\% core accesses but does not restore the lost performance with only 3\% IPC gain over ARP-NB. 

As discussed in Section \ref{sec:keyInsight2}, we observe that \tikz[baseline=(X.base)] \protect\node (X) [draw, circle, fill=white, text=black, inner sep=1.5pt] {\small{5}}; ARP-CAS shows accelerator BR closely following the core BR, indicating that shared reuse predictors cause access pattern interference. \tikz[baseline=(X.base)] \protect\node (X) [draw, circle, fill=white, text=black, inner sep=1.5pt] {\small{6}}; ARP-CS-AS, on the other hand, minimizes the interference between them, shows a higher accelerator BR, and recovers the lost IPC but leads to deadline misses due to deadline unawareness.

We also observe that \tikz[baseline=(X.base)] \protect\node (X) [draw, circle, fill=white, text=black, inner sep=1.5pt] {\small{7}}; compared to ARP-AS, ARP-AL achieves a higher accelerator BR with a 40\% drop in the DMR without impacting IPC, \tikz[baseline=(X.base)] \protect\node (X) [draw, circle, fill=white, text=black, inner sep=1.5pt] {\small{8}}; ARP-AS-D meets the deadline, but a steep drop in the accelerator BR due to deadline-awareness causes IPC degradation. With ARP-AL-D, not only is the deadline met, but the IPC is also restored, with a 1\% speedup over FIFO-NB. 
Therefore, LERN performs better at predicting the accelerator's reuse patterns.

\tikz[baseline=(X.base)] \protect\node (X) [draw, circle, fill=white, text=black, inner sep=1.5pt] {\small{9}}; ARP-CS-AS-D being deadline-aware meets the deadline with a 2\% drop in the IPC compared to FIFO-NB. 
HyDRA (ARP-CS-AL-D), on the other hand, not only meets the accelerator's deadline but also achieves a 4\% IPC speedup with 50\% DMR reduction over FIFO-NB and a 6.1\% IPC speedup over the ARP-CS-AS-D policy. 
Hence, HyDRA can achieve an efficient balance between deadline and reuse awareness. 
For brevity, we choose the following baselines for a detailed performance comparison across different accelerator configurations: FIFO-NB, ARP-NB, ARP-AS-D, and ARP-CS-AS-D.

\subsubsection{Performance breakdown of HyDRA on Config-1 across different workload mixes.}

Figure \ref{fig:config1-main} shows the performance comparison of HyDRA against the deadline-aware SHIP-driven bypass baselines. On workload \emph{mix1}, which simulates CI applications on all cores with a very low LLC access rate, the performance of all the policies is within 1\% of each other. 
Workloads \emph{mix2} - \emph{mix5} simulate LI benchmarks on all cores, posing high LLC contention to the accelerator. 
HyDRA achieves IPC speedups of up to 8\% and 11\% over FIFO-NB and ARP-CS-AS-D, respectively. For other workload mixes with varying combinations of CI, LI, and MI applications, HyDRA achieves significantly better performance than ARP-CS-AS-D whenever the cache contention is high.

\subsubsection{{Detailed analysis of accelerator's epoch-wise LLC access rate with HyDRA on Config-1.}}
{Figure \ref{fig:mix4AccessRate} shows the epoch-wise LLC access rate variation of the accelerator Config-1 with \emph{mix4} along with the evolution of the per-epoch progress requirement. The epoch-wise access completion rate is non-uniform with large spikes, deep troughs, and long regions of sustained high and low access rates with abrupt transitions based on the progress evaluation. We observe that the access rate changes dramatically and significantly over epochs. At the beginning of each frame, HyDRA alternates between low- and high-access phases by varying the bypass aggressiveness of the accelerator as progress is made. However, towards the end, we observe that the access rate drops sharply as the per-epoch progress requirement decreases. 
We also observe that in some epochs, actual accesses completed are below the required accesses. In other epochs, actual accesses greatly exceed the requirement. This shows that the access rate is highly non-uniform throughout the execution.}

\begin{figure*}[htbp]
    \centering
    \includegraphics[width=\linewidth]{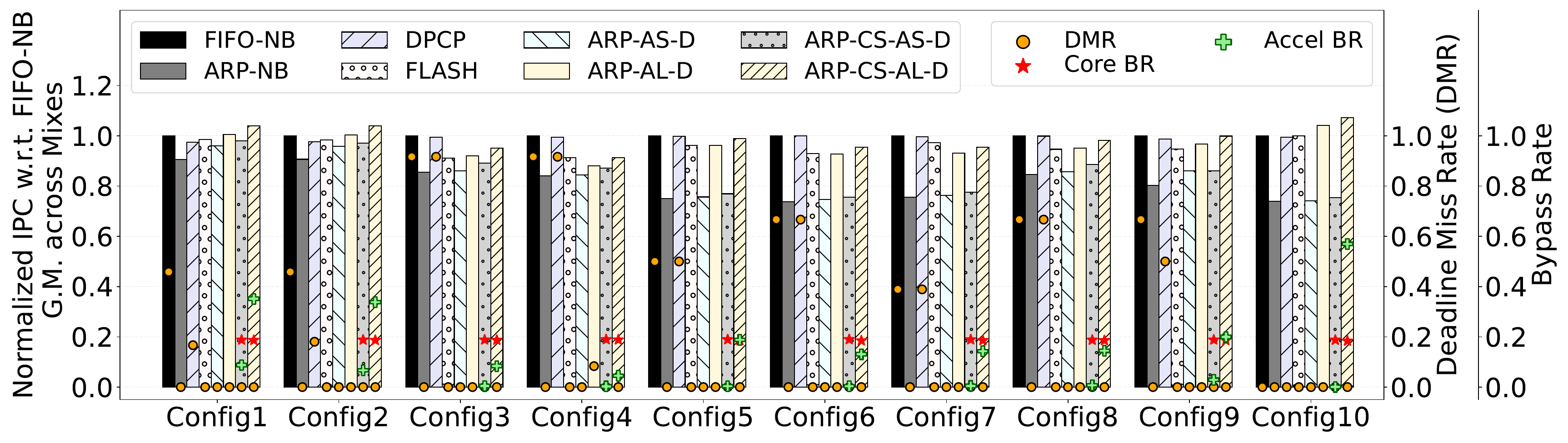}
    \caption{Performance Evaluation of HyDRA across Accelerator Configurations Config-1 to Config-10. Deadline: 10 IPS.}
    \label{fig:AllConfigPerf}
\end{figure*}
\begin{figure*}[htbp]
    \centering
    \includegraphics[width=\linewidth]{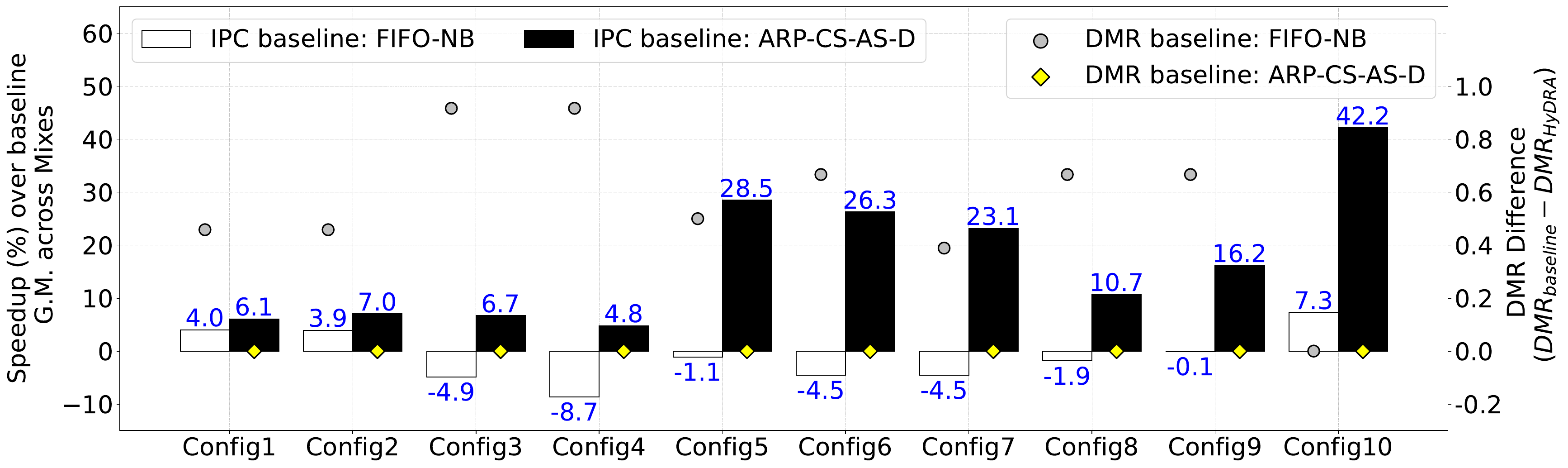}    
    \caption{{Performance Comparison of HyDRA with FIFO-NB and ARP-CS-AS-D. Deadline: 10 IPS.}}
    \label{fig:hydravsbaseline}
\end{figure*}

\subsubsection{Performance comparison of HyDRA across different accelerator configurations with varying memory intensity and access patterns.}

Figure \ref{fig:AllConfigPerf} shows the mean performance of HyDRA over all the mixes compared to the baseline policies for all the accelerator configurations, Config-1 to Config-10 (Table \ref{tab:hwaconfig}). 
Figure \ref{fig:hydravsbaseline} compares HyDRA with the default LLC management policy, FIFO-NB, and with the ARP-CS-AS-D policy as it achieves the same DMR and hence, presents a fair comparison. 

Config-2 is similar to Config-1 but simulates the model on the accelerator with WS dataflow. 
As shown in Figures \ref{fig:AllConfigPerf} and \ref{fig:hydravsbaseline}, the performance of all the policies is relatively similar to Config-1, as the large on-chip SRAMs hide the access pattern difference between the two dataflows.
Config-3 and Config-4 have smaller on-chip memory and present higher memory intensity, partially due to repetitive accesses at the shared cache, and hence, pose higher contention with a higher cache hit rate. 
LERN achieves up to 10\% accelerator BR in contrast to an almost negligible accelerator BR achieved by SHIP (due to high hit rate) with deadline awareness, and shows better performance. Hence, HyDRA achieves an IPC speedup of up to 6.7\% over ARP-CS-AS-D policy and reduces the DMR by more than 90\% compared to FIFO-NB. 
Config-5 and Config-6 simulate the GoogleNet model on an edge TPU equivalent accelerator with OS and WS dataflow, respectively. 
While SHIP cannot detect the dataflow-dependent variations in the memory access pattern, LERN captures the variations and shows a slightly lower accelerator BR with WS than OS dataflow. Hence, HyDRA shows up to 28\% IPC improvement over ARP-CS-AS-D and reduces the DMR by more than 50\% over FIFO-NB. 
Config-7 simulates MobileNet on an edge TPU equivalent accelerator and shows a similar access pattern and performance as Config-6. 
Config-10 simulates the AlphaGoZero network model on an edge-TPU equivalent accelerator and shows a very low memory intensity on the shared cache. 
HyDRA meets the deadline and provides an IPC speedup of 7.3\% and 42\% over FIFO-NB, and ARP-CS-AS-D, respectively. 
Config-8 and Config-9 simulate Deep Speech2 and Faster R-CNN on edge TPU and serve-grade TPU equivalent accelerators and show memory intensity between Config-10 (very low) and Config-4 (very high). HyDRA achieves more than 60\% reduction in DMR over FIFO-NB and IPC improvement of up to 16.2\% over the ARP-CS-AS-D policy.

\subsubsection{Contribution of reuse-aware core bypass to the IPC speedup.}
We compare the performance of HyDRA with core bypass (ARP-CS-AL-D) and HyDRA's adaptation without core bypass (ARP-AL-D) to quantify the IPC speedup achieved with core bypass at the shared cache.
Across all the accelerator configurations, core bypass adds up to 2.6\% -- 4\% IPC speedup and up to 8\% DMR reduction (on Config-4). 
Hence, adding reuse-aware core bypass using any state-of-the-art predictor further alleviates space and bandwidth contention and improves throughput. 

\begin{figure*}[htbp]
    \centering
    \includegraphics[width=\linewidth]{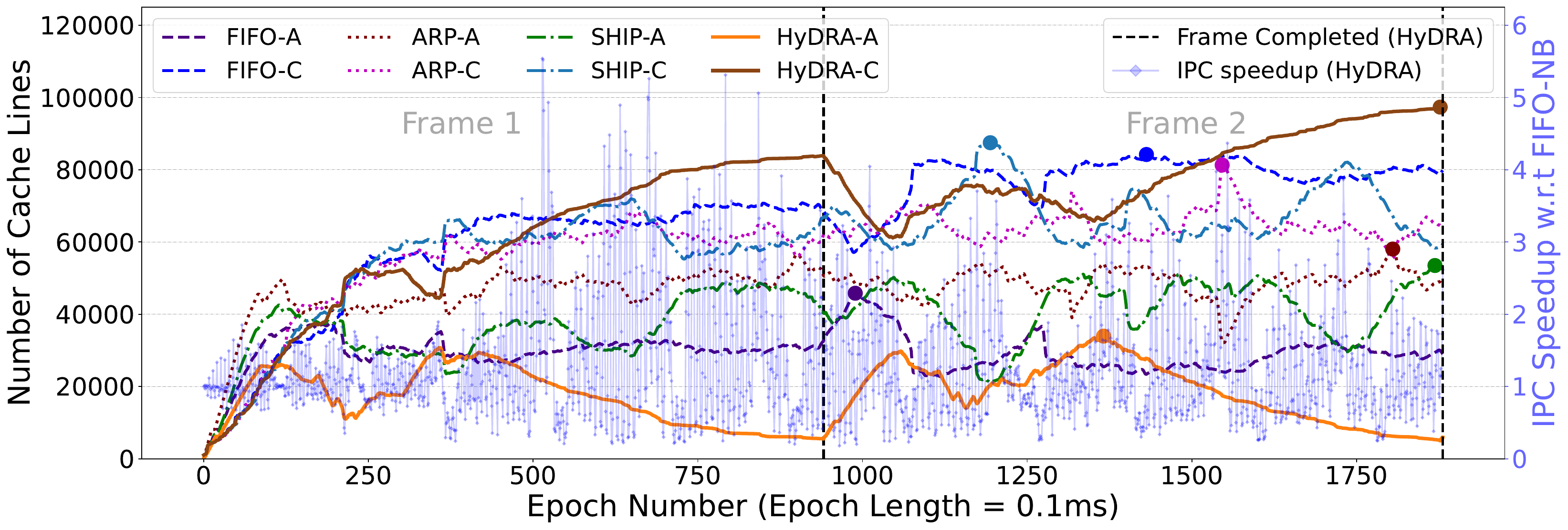}
    \caption{Comparison of the shared cache space occupied by all cores (-C) and the accelerator (-A) during the execution with different cache management policies. Accelerator Configuration: Config-1, \emph{mix3}. Markers on the plot denote the maximum value across frames. Total shared cache lines = $131072$ ($8$ MB, $64$ B blocks). Deadline: 10 IPS.}
    \label{fig:config1Mix3CacheSpace}
\end{figure*}

\subsubsection{{Comparison of cache occupancy across different cache management policies.}}
{HyDRA uses deadline-aware cache bypass to reclaim both cache space and bandwidth from the accelerator, improving the performance of cores. 
Figure \ref{fig:config1Mix3CacheSpace} shows the comparison of the number of cache lines occupied by all processor cores together (denoted with \emph{-C}) and the accelerator (denoted with \emph{-A}) across different cache management policies, such as FIFO-NB, ARP-NB, ARP-CS-AS-D (abbreviated as SHIP), and ARP-CS-AL-D (abbreviated as HyDRA), during the execution of the accelerator with configuration Config-1 and SPEC benchmarks \emph{mix3}. It also shows the maximum number of cache lines occupied during the execution in bold markers.}

{Under FIFO-NB, cores occupy most of the cache while the accelerator experiences higher contention, leading to higher deadline miss rates. ARP-NB prioritizes the accelerator, increasing its cache occupancy by up to 27\% but reducing space available to cores. ARP-CS-AS-D moderates this through deadline and SHIP-driven reuse-aware bypass, slightly reducing accelerator occupancy by up to 8\% compared to ARP-NB. In contrast, HyDRA dynamically adjusts the bypass based on deadline progress, significantly reducing accelerator cache usage by up to 36.5\% and 25.8\%, while increasing cores' cache occupancy by up to 11.2\% and 15.5\% compared to ARP-CS-AS-D and FIFO-NB, respectively. As a result, HyDRA achieves higher cache space allocation for cores, meets accelerator deadlines, and improves IPC on \emph{mix3} by about 2\% while eliminating deadline misses, demonstrating effective redistribution of cache resources between CPUs and the accelerator.}

\subsection{{Performance Comparison of HyDRA with prior cache management policies DPCP and FLASH.}}
{We implement DPCP~\cite{DPCP} by statically allocating 1 cache way (out of 16 ways) to the accelerator and enabling a stride prefetcher at the LLC for accelerator accesses. All accesses (both demand and prefetch) from the accelerator are restricted to a single cache way across all sets. 
As shown in Figure \ref{fig:AllConfigPerf}, while prefetching reduces DMR for some configurations (Configs 1, 2, and 9), static partitioning limits cores' performance, resulting in mean performance comparable to or worse than FIFO-NB. As DPCP is not QoS-aware, it misses deadlines across most configurations and cannot match HyDRA's performance, highlighting the advantage of HyDRA’s deadline- and reuse-aware cache management.}

{
As discussed in Section \ref{sec:relwork}, FLASH~\cite{FLASH} dynamically partitions shared cache bandwidth between cores and accelerators based on accelerator progress, but assumes all accelerator requests must be served by the shared cache, which can lead to inefficient cache utilization. 
HyDRA investigates complementary cache bypassing techniques to manage cache interference between accelerators and cores and to leverage deadline and reuse awareness to determine the selective cacheability of accelerator accesses in the shared cache, thereby regulating both space and bandwidth contention. 
As shown in Figure \ref{fig:AllConfigPerf}, while both approaches aim to improve system performance while meeting accelerator deadlines, HyDRA achieves better cache utilization and up to 7.3\% higher IPC than FLASH. The main contributor to the performance improvement is the efficient utilization of cache space and bandwidth via reuse- and deadline-aware cache bypass.}

\begin{figure*}[htbp]
    \centering
    \begin{subfigure}[b]{\linewidth}
        \centering
        \includegraphics[width=\linewidth]{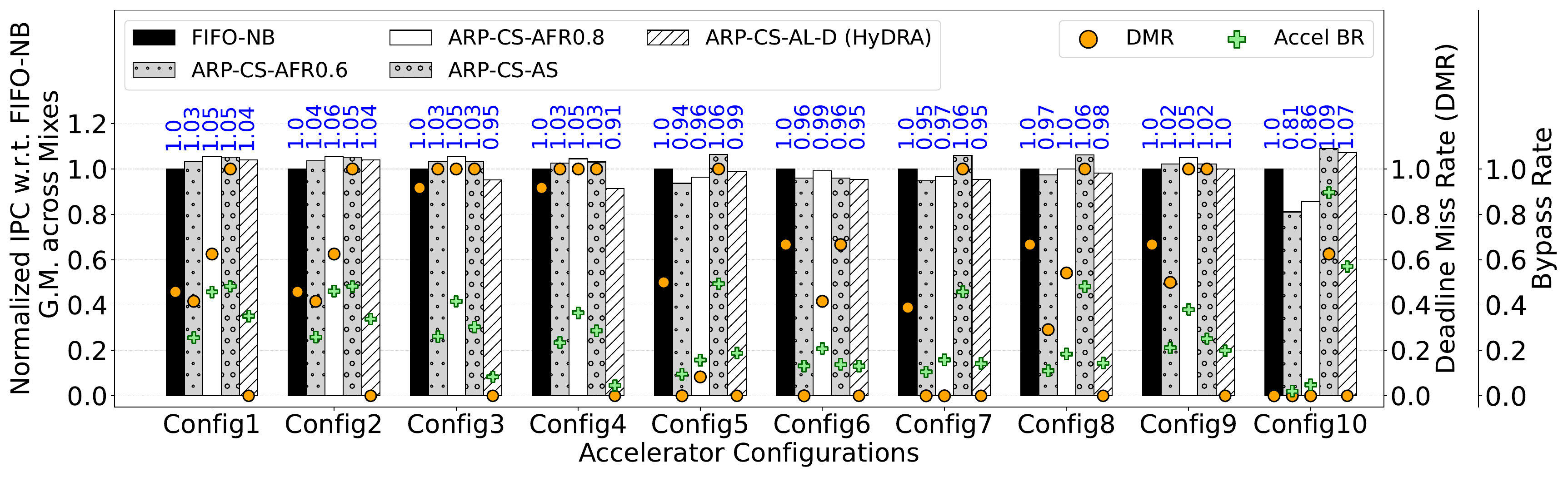}
        \caption{Performance achieved by deadline-agnostic SHIP-driven bypass policy and policies that choose accelerator bypass contenders randomly with 60\% (AFR0.6) and 80\% (AFR0.8) probability.}
        \label{fig:FRBypassComp}
    \end{subfigure}
    \par
    \begin{subfigure}[b]{\linewidth}
        \centering
        \includegraphics[width=\linewidth]{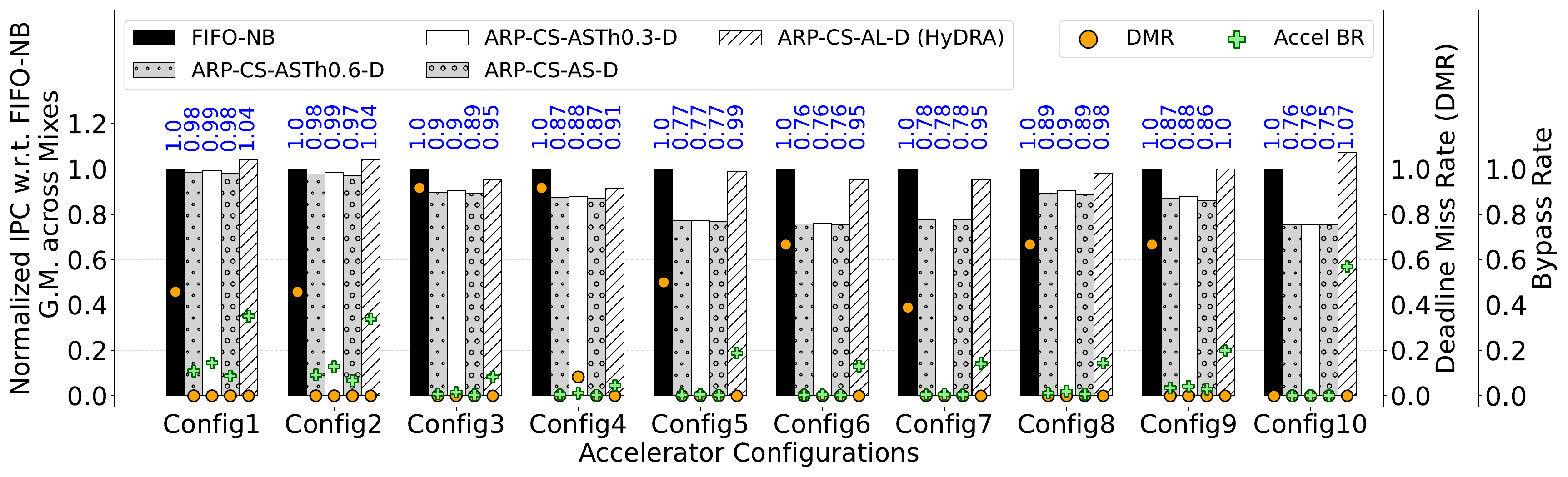}
        \caption{Performance achieved by HyDRA over deadline-aware SHIP-driven bypass and FIFO-NB policies. Accelerator bypass starts after 30\% (ASTh0.3), 60\% (ASTh0.6), and 100\% (AS) of the required memory accesses are completed in an epoch.}
        \label{fig:ShipWithDComp}
    \end{subfigure}
    \caption{Performance achieved by policies with similar accelerator bypass rates as HyDRA. Deadline: 10 IPS.}
    \label{fig:similarBypRate}
\end{figure*}

\subsection{{Comparison with ARP-CS-AS (deadline-agnostic SHIP-driven core and accelerator bypass)}}
{In Figure \ref{fig:config1-all}, we compare HyDRA's performance on Config-1 (mean across all mixes) with several deadline-agnostic, deadline-aware, and reuse-aware cache management policies, including ARP-CS-AS and ARP-CS-AS-D. We observe that the deadline-agnostic (no knowledge of the deadline, as in HyDRA) SHIP-driven bypass policy, ARP-CS-AS, achieves similar accelerator bypass rates as HyDRA but with a high DMR of 1. We evaluate the performance of the policy ARP-CS-AS on other accelerator configurations. As shown in Figure \ref{fig:FRBypassComp}, ARP-CS-AS does not meet the accelerator deadline as it does not have the knowledge of the accelerator's dynamic progress. For Config-1 -- Config-4, ARP-CS-AS achieves 10\%--20\% higher bypass for accelerator accesses, with 1\%--12\% and 100\% higher IPC and DMR, respectively, than HyDRA. On Config-6, ARP-CS-AS achieves an almost identical accelerator bypass rate, but with more than 60\% DMR.}

\subsection{{Performance Evaluation of deadline-agnostic probabilistic accelerator bypass policy}}
{We design a probabilistic bypass policy, ARP-CS-AFR$p$, that randomly selects bypass contenders for accelerator accesses with probability $p$. Figure \ref{fig:FRBypassComp} demonstrates the performance achieved by policies ARP-CS-AFR0.6 ($p=0.6$) and ARP-CS-AFR0.8 ($p=0.8$). However, if a read memory access is a bypass contender but is already present in the cache, it is not bypassed. We observe that ARP-CS-AFR0.6 achieves accelerator bypass rates similar to or higher than HyDRA's, but it does not meet the accelerator deadline for all configurations except Configs 5, 6, and 7. On Configs 6 and 7, ARP-CS-FR0.6 shows similar performance to HyDRA. 
Similarly, ARP-CS-AFR0.8 also shows non-zero DMR for all accelerator configurations except Configs 7 and 10, with 2\% IPC improvement on Config-7 and 21\% IPC degradation on Config-10 compared to HyDRA. 
Therefore, even though other policies can achieve bypass rates similar to HyDRA, without HyDRA's knowledge, accelerator deadlines cannot be met.}

\subsection{{Detailed Comparison with deadline-aware SHIP-driven accelerator bypass with different bypass aggressiveness}}
{We implemented a deadline-aware extension to SHIP-driven accelerator bypass to compare the performance improvements achieved by deadline-aware LERN-driven bypass. Deadline-aware SHIP-driven bypass is implemented as discussed in Section \ref{sec:deadlineAwareness}. The accelerator bypass begins once the epoch-level progress requirement is met. Since meeting the deadline is the primary objective, the ARP-CS-AS-D policy is insertion-heavy at the beginning of the epoch and becomes bypass-heavy towards the end when the epoch-level progress requirement is met. 
We create ARP-CS-ASTh$t$-D, which starts bypassing the accelerator accesses (according to the SHIP predictor) after $t$ times the required memory accesses have been completed. Figure \ref{fig:ShipWithDComp} discusses the performance achieved by two variations with $t=0.3 \mbox{ and } 0.6$. We observe that with ARP-CS-ASTh0.6-D, the accelerator bypass rate slightly increases, with no impact on performance compared to ARP-CS-AS-D. The policy ARP-CS-ASTh0.3-D, on the other hand, begins bypass as soon as 30\% of the progress requirement is met in the epoch. We observe that for Config-4, this policy increases the DMR by almost 10\% with a 1\% IPC improvement, and therefore, cannot guarantee meeting deadlines. We also observe that the performance improvements are not comparable to HyDRA.}

\subsection{{Evaluation across different system architectures.}}
{The system architecture used to evaluate HyDRA targets heterogeneous edge systems such as mobile, automotive, networking, and edge-IoT SoCs, where cache sizes are constrained by area and power. In these systems, 256 KB private L2 caches per core are common, particularly for efficiency and accelerator-adjacent cores, as seen in commercial platforms such as Qualcomm Snapdragon/Kryo processors~\cite{Kryo585, QSoCBreakdown, QualRB5, QCS8250}, Huawei Kirin SoCs~\cite{kirin9000}, NVIDIA Jetson Orin~\cite{orinSoC}, and MediaTek Dimensity devices~\cite{Mediatek8400}. However, commercial systems also include configurations with larger private caches and varying shared cache capacities, as well as varying memory systems. Most modern systems also enable hardware prefetchers in the private and shared caches. To understand HyDRA's sensitivity to these architectural parameters, we evaluate its performance across different L2 sizes (e.g., 512 KB), shared LLC capacities (1--16 MB), DDR memory models (DDR4, DDR5), and with prefetchers enabled at the private caches. This analysis helps determine whether HyDRA’s benefits persist under varying cache hierarchies and memory system characteristics representative of real-world edge platforms.}

\subsubsection{{Performance Evaluation with 512KB private L2 cache.}}
{With a 512KB private L2 cache, FIFO-NB observes up to 63\% fewer LLC accesses (on Config-1), significantly reducing interference between cores and the accelerator and allowing the accelerator to meet deadlines in most cases, except the more memory-intensive Configs 3 and 4. HyDRA improves IPC by up to 3\% over FIFO-NB and 6\% over ARP-CS-AS-D, and on Configs 3 and 4, reduces DMR by up to 35\% (FIFO-NB) while incurring only 1\% IPC degradation.}

\begin{figure}[t!]
    \centering
    \begin{subfigure}[b]{\linewidth}
        \centering
        \includegraphics[width=\linewidth]{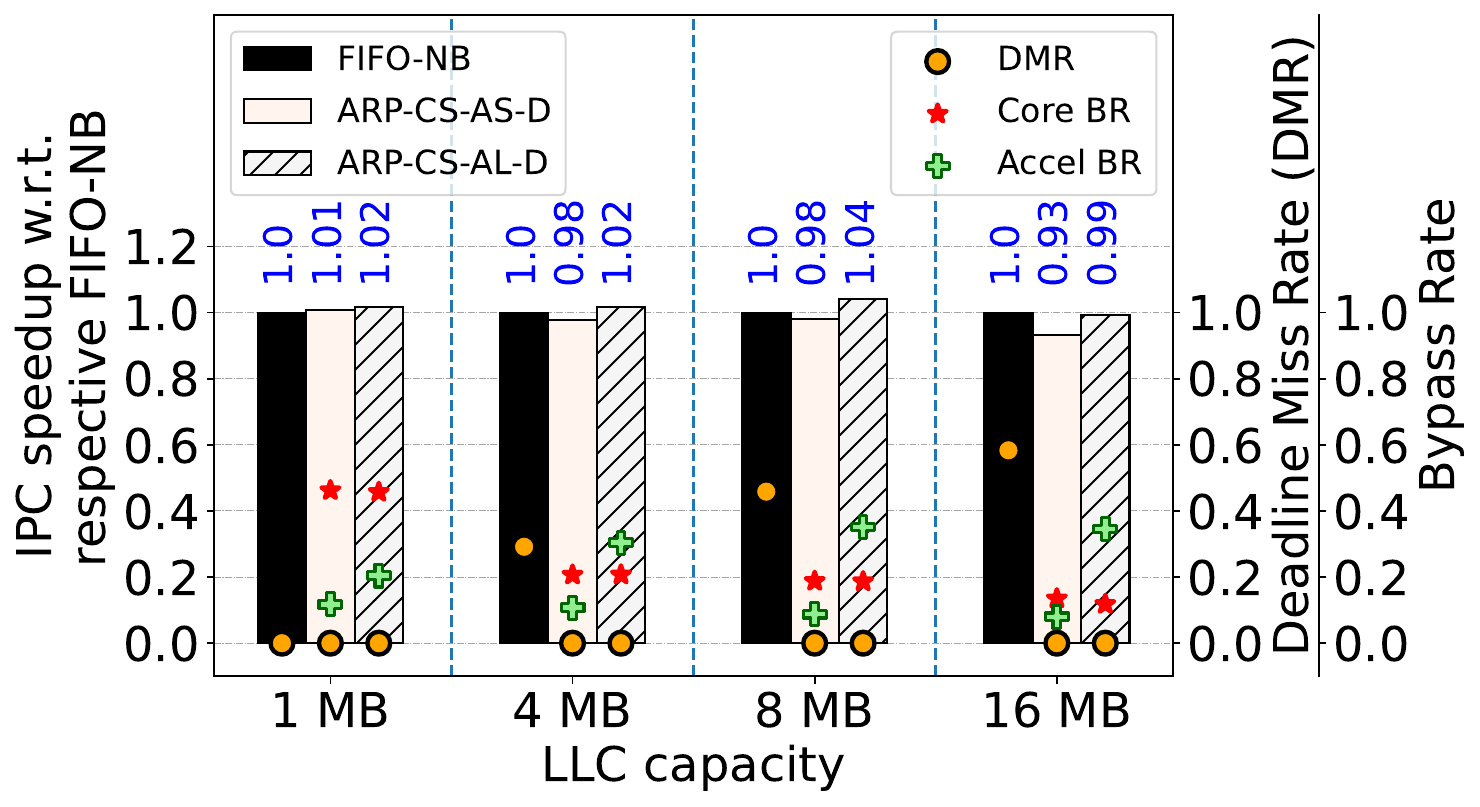}
        \caption{Config-1. Deadline: 10 IPS.}
        \label{fig:C1-L3SizeComp}
    \end{subfigure}
    \par
    \begin{subfigure}[b]{\linewidth}
        \centering
        \includegraphics[width=\linewidth]{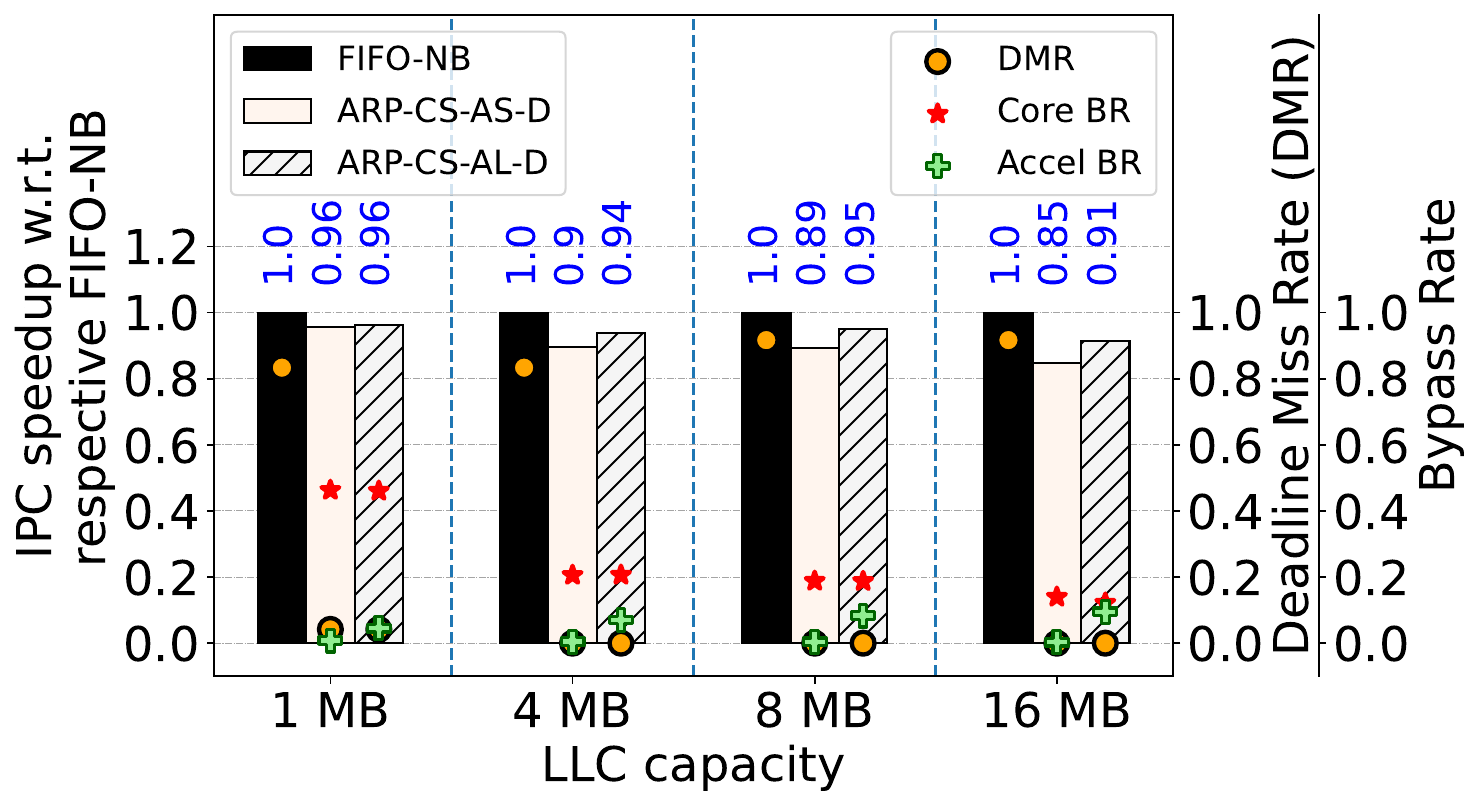}
        \caption{Config-3. Deadline: 10 IPS.}
        \label{fig:C3-L3SizeComp}
    \end{subfigure}
    \caption{Performance Evaluation with varying LLC capacity. The performance of HyDRA across LLC sizes is normalized to its respective FIFO-NB performance. }
    \label{fig:C1C3-L3SizeComp}
\end{figure}

\subsubsection{{Performance Evaluation with varying LLC capacity.}}
{We also evaluate HyDRA across shared LLC sizes from 1 MB to 16 MB on Config-1.  Figures \ref{fig:C1-L3SizeComp} and \ref{fig:C3-L3SizeComp} show the performance of different cache management policies with varying cache capacity on accelerator configurations Config-1 and Config-3, respectively, with a 10 IPS deadline. For each LLC size, we compare the performance of HyDRA and ARP-CS-AS-D with that of FIFO-NB.}

{
With a 1 MB cache, high core-to-core interference increases cache miss rates and makes processor workloads memory-bound. 
As a result, the accelerator accesses get higher LLC bandwidth, and the FIFO-NB policy also meets the deadline. We also observe that cache bypass policies bypass more than 40\% of core accesses, as reuse is low due to interference; however, this does not translate into higher speedups as off-chip memory becomes the bottleneck.
HyDRA provides only modest gains of up to 2\% over the FIFO-NB policy. 
In accelerator configurations Configs 3 and 4, we also observed that with 1 MB LLC cache, the accelerator is unable to meet the deadline with the ARP-NB policy (priority arbitration at the LLC), as these configurations exhibit high memory intensity and higher contention at the shared cache. Therefore, a very small shared cache for an EdgeTPU-grade accelerator might not be sufficient to meet the deadlines. 
}

{
As the cache size increases, interference decreases, and the available cache space per core increases, leading to higher cache hit rates per core. Accelerator accesses take longer to be serviced by the cache. 
With 4 MB and 8 MB shared cache, HyDRA consistently meets deadlines while improving IPC (2--4\%) and significantly reducing DMR (up to 45.8\%) compared to FIFO-NB. With a 16 MB LLC, HyDRA reduces deadline miss rate by up to 60\% with only 1\% performance loss. 
As shown in Figure \ref{fig:C3-L3SizeComp}, we observe similar benefits over the ARP-CS-AS-D policy on Config-3 as well. Compared to the FIFO-NB policy, HyDRA reduces the DMR by upto 90\% on Config-3 with up to 9\% degradation in performance. 
}

{
Overall, HyDRA ensures meeting deadlines across cache capacities while maintaining performance comparable to or better than FIFO-NB.
}

\subsubsection{{Performance Evaluation with different memory models.}}
{The simulated system configuration uses a DDR3\_1600\_8x8 (as shown in Table \ref{tab:systemconfig}) memory model with DDR3 memory running at 1600MHz, with 2 ranks, 8 banks/rank, and a 64-bit interface. Each bank is 512KB in size with an 8-bit interface. With a burst size of 8, this memory provides 64B in one memory access. 
We evaluate HyDRA on two other memory models, DDR4\_2400\_8x8 and LPDDR5\_5500\_1x16\_BG\_BL16. The DDR4\_2400\_8x8 memory model has 2 ranks, 8 banks per rank, and a 64-bit interface. Each bank is 1 GB in size. The LPDDR5\_5500\_1x16\_BG\_BL16 memory model simulates LPDDR5 with 4 bank groups of 4 banks each at 5500 MHz and a 16-bit interface. BL16 denotes a burst size of 16 (2 bytes each). So a single access provides 32B data as opposed to 64B provided by DDR3 memory. 
DDR3 and DDR4 typically have lower access latency in cycles. Since LPDDR5 provides 32B per access, each cache line access takes longer to complete than in DDR3 and DDR4. LPDDR5 achieves its speed through complex pipelining and bank group parallelism.}

\begin{figure}[t!]
    \centering
    \includegraphics[width=\linewidth]{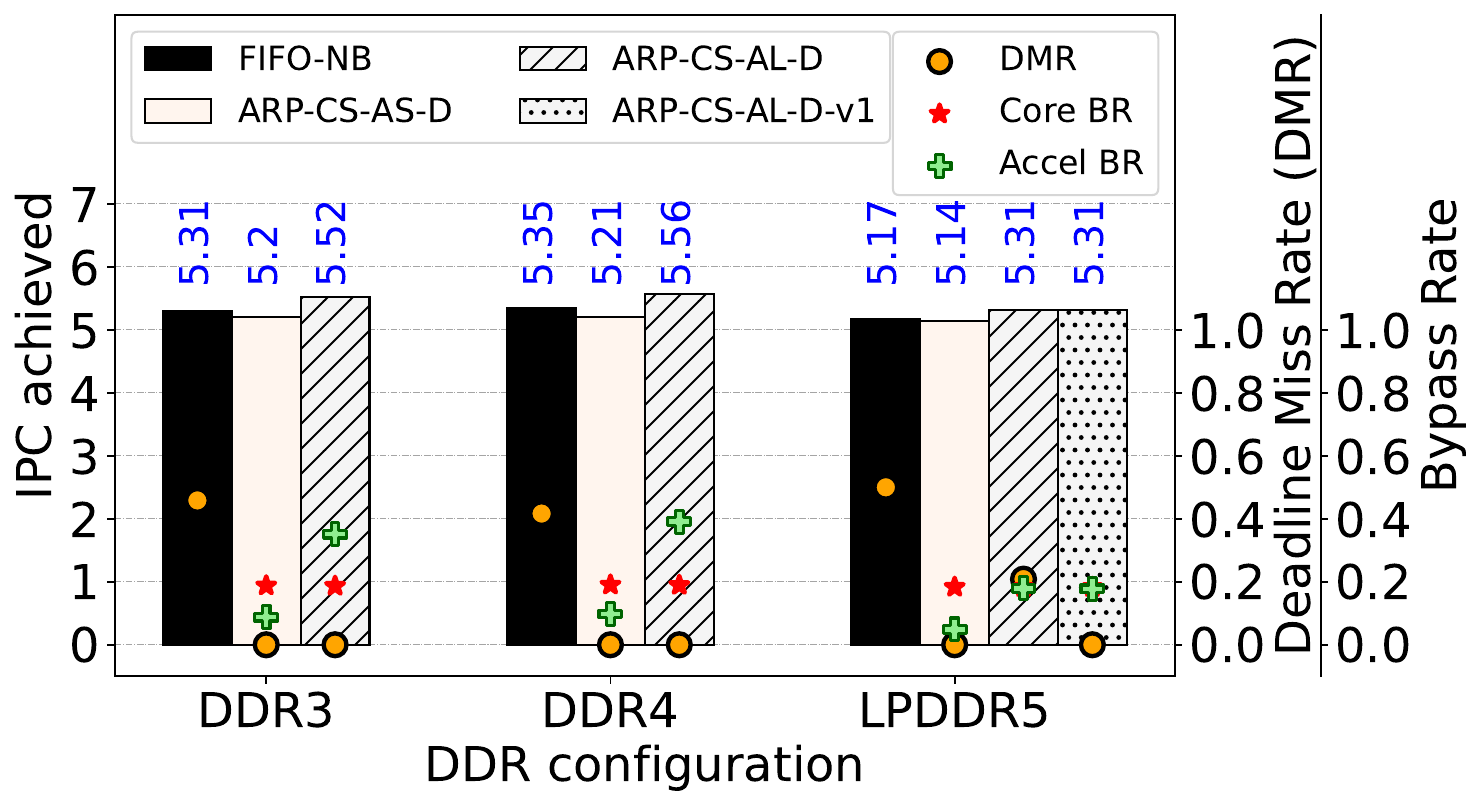}
    \caption{Performance evaluation of HyDRA with different DDR models on Config-1 (mean performance across mixes). Deadline: 10 IPS.}
    \label{fig:C1-ddrComp}
\end{figure}

{
As shown in Figure \ref{fig:C1-ddrComp}, moving to DDR4 yields nearly identical performance to DDR3, with HyDRA achieving 4\% IPC improvement and over 40\% DMR reduction compared to FIFO-NB. In contrast, LPDDR5’s smaller burst size and higher effective latency increase DMR for FIFO-NB and initially cause HyDRA to miss deadlines, as its parameters are tuned for DDR3. We observe a 4.2\% and 9.4\% increase in the DMR of the FIFO-NB policy over DDR3 and DDR4, respectively.  
We evaluate a variation of HyDRA, ARP-CS-AL-D-v1 (HyDRA-v1), with parameters tuned for the DDR5 model. We increase the safety margins for recovering from the impact of high memory contention due to longer raw latencies and use $\mbox{margin}_{\mbox{\scriptsize high}}=10\%$ and $\mbox{margin}_{\mbox{\scriptsize low}}=2\%$. The other parameters remain the same as ARP-CS-AL-D (HyDRA). We observe that the accelerator bypass rate decreases by 0.1\% compared to HyDRA, and HyDRA-v1 meets deadlines without performance loss. Overall, HyDRA’s performance gains remain robust across memory models, though higher raw latencies may require minor fine-tuning to maintain deadline guarantees.}

\subsubsection{{Performance Evaluation with Stride Prefetchers enabled at the private L2 caches.}}
To understand the robustness of our approach, we evaluate a system configuration with L2 prefetchers enabled and observe that the overall performance trends remain similar.  
When L2 Stride prefetchers are enabled, they increase interference with the accelerator's demand traffic. On Config-1, the FIFO-NB policy shows an increase in DMR by up to 20\% and an IPC speedup of up to 3\% with prefetchers enabled. 
Both HyDRA and the FIFO-NB baseline experience higher LLC pressure, yet the relative trends remain consistent: HyDRA meets the deadline and reduces the DMR by more than 60\% while improving the IPC by up to 2\%. This indicates that the proposed deadline- and demand-aware LLC regulation is robust even in the presence of prefetch traffic.

\begin{figure}[t!]
    \centering
    \includegraphics[width=\linewidth]{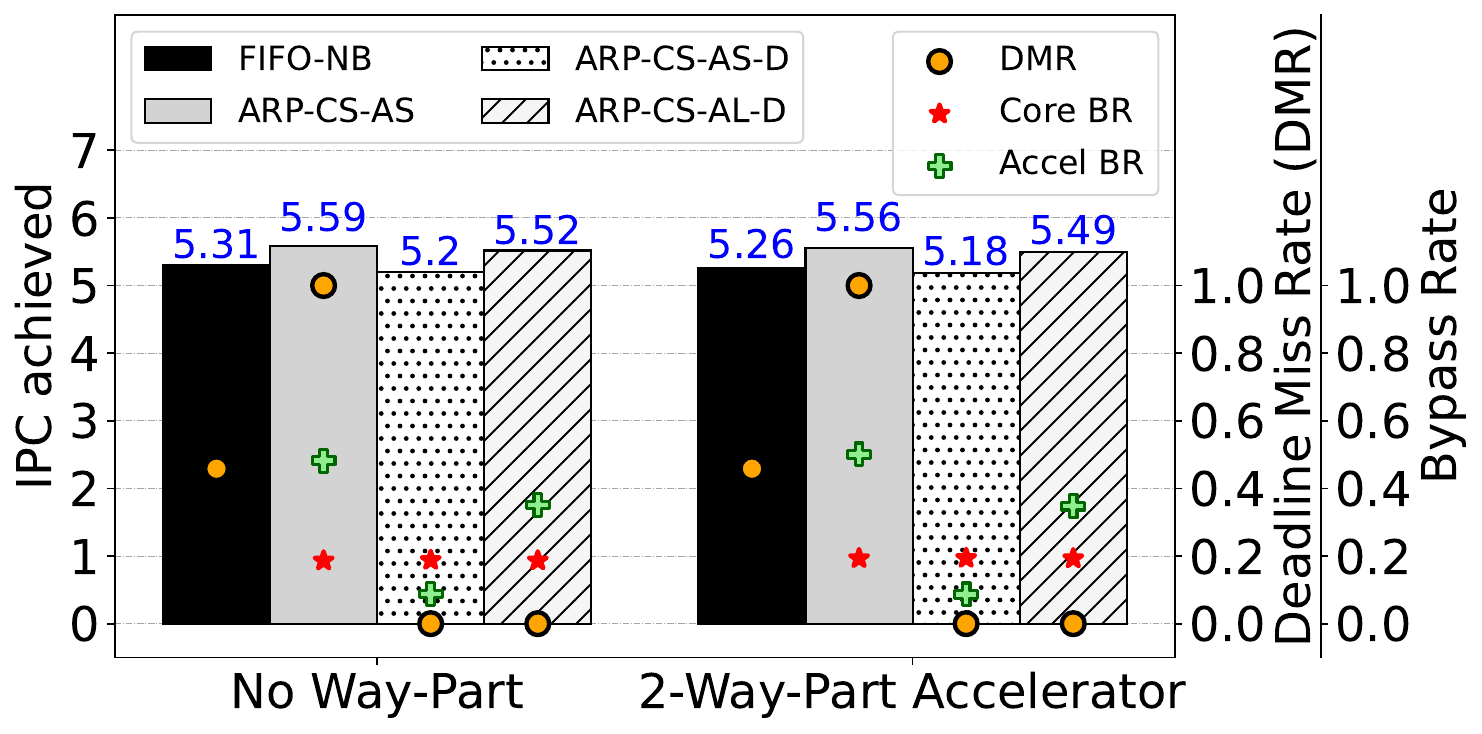}
    \caption{Performance evaluation with 2-way cache partitioning for Accelerator on Config-1. Deadline: 10 IPS. }
    \label{fig:2wayConfig1}
\end{figure}

\subsection{{Performance Evaluation with Cache Way Partitioning.}}
{We implement cache-way partitioning in \emph{gem5} to isolate cache space contention between processor cores and accelerators. We allocate 2 ways of the shared cache to the accelerator, and the remaining 14 ways are shared between processor cores, eliminating direct cache-space interference. Figure \ref{fig:2wayConfig1} shows the performance achieved by different cache management policies, including HyDRA, on accelerator Config-1 with and without way partitioning. We observe that with 2 ways reserved for the accelerator, FIFO-NB policy shows a 1\% drop in IPC achieved with no change in the DMR. Hence, way partitioning might not be sufficient to meet accelerator deadlines without a significant drop in cores' performance (with a higher number of ways reserved for the accelerator).
HyDRA with way partitioning, on the other hand, shows a 0.5\% reduction in speedup compared to HyDRA with no way partitioning, with a slightly lower accelerator bypass rate. Hence, way partitioning in shared caches slightly reduces cores' performance. However, the impact on IPC is lower with the HyDRA cache management policy. Additionally, HyDRA achieves 4.3\% performance improvement over the FIFO-NB policy with 2-way partitioning, compared to 3.9\% without 2-way partitioning.}

\begin{figure*}[htbp]
    \centering
    \includegraphics[width=\linewidth]{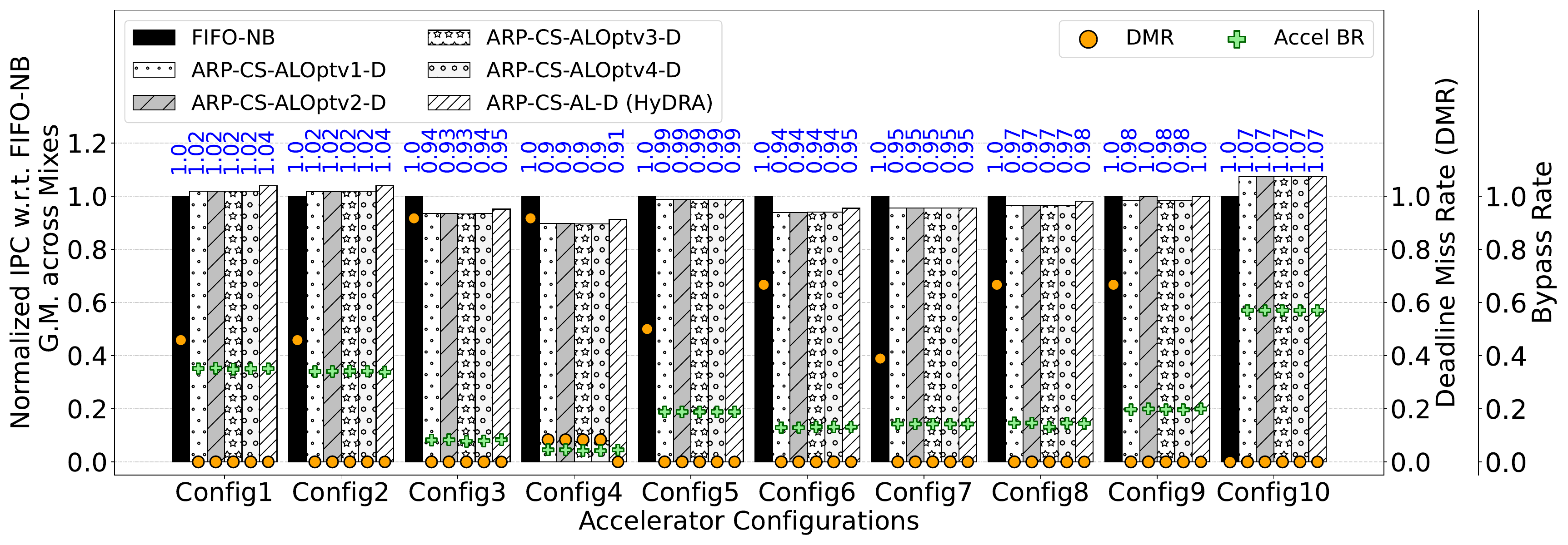}
    \caption{Performance evaluation of HyDRA with different LERN predictor table entries. LERN is trained on hashed addresses. Deadline: 10 IPS. }
    \label{fig:LernOpt}
\end{figure*}

\begin{figure*}[htbp]
    \centering
    \includegraphics[width=\linewidth]{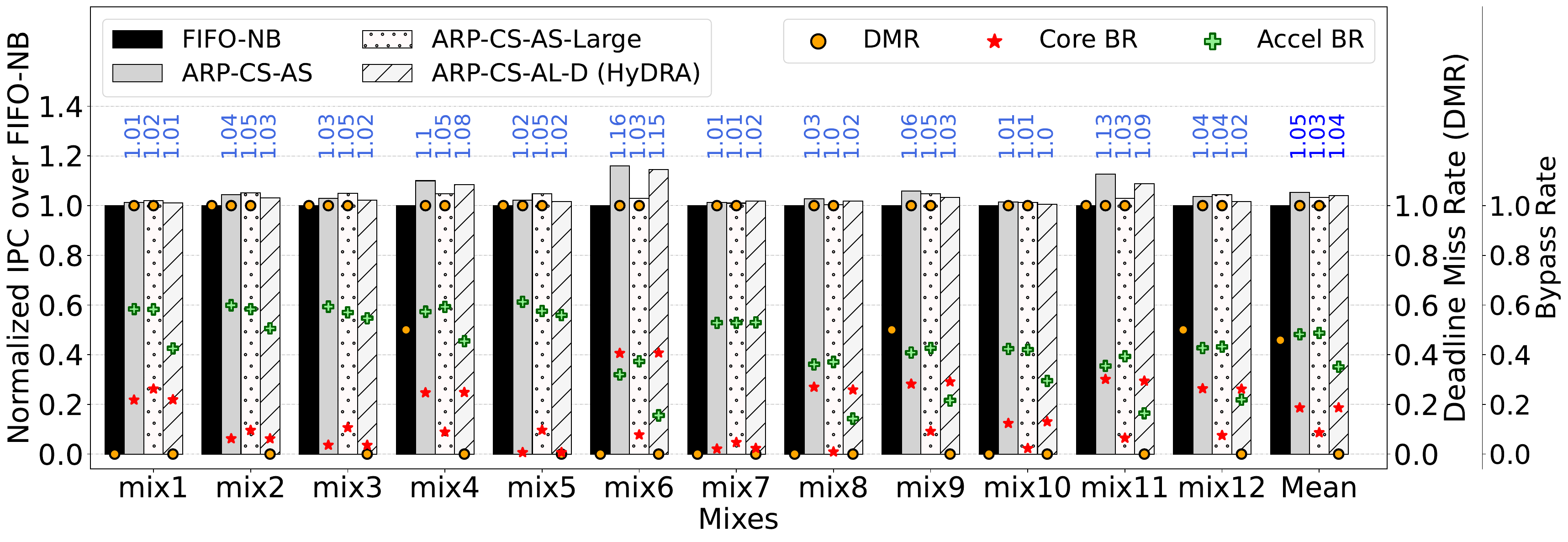}
    \caption{Performance evaluation of HyDRA over FIFO-NB and SHIP-driven bypass with different SHIP predictor table size on Config-1. Deadline: 10 IPS.}
    \label{fig:shipLarge}
\end{figure*}

\subsection{{Performance Evaluation with hardware-optimized L-RPT}}
\label{sec:lrptOpt}
{When the required entries per layer exceed the hardware L-RPT capacity (512K entries in the current evaluated system), LERN predictor training is performed on block addresses hashed into the reduced table space so the predictor learns under the same aliasing conditions as the hardware and internalizes the runtime conflicts to adapt its cluster assignments accordingly. This prevents a mismatch between offline learning and hardware inference, which could otherwise degrade prediction accuracy.}

{To optimize the hardware overhead of L-RPT, we conduct a design-space exploration by evaluating reduced table sizes (25\% and 50\% of the original 512K-entry configuration) and analyzing their performance impact. 
We evaluate L-RPT designs with 128K and 256K entries using two hashing methods: Bitmask and SplitMix32~\cite{splitmix}. With 5 bits of data per entry, the size of these L-RPT designs with 128K and 256K entries is 80KB and 160KB, respectively, as opposed to the 320KB size of L-RPT with 512K entries. Figure \ref{fig:LernOpt} shows the performance achieved by HyDRA with the LERN predictor trained across different table sizes and address hashes. We evaluate 4 different versions of L-RPT size and hashing methodology:
\begin{itemize}
    \item LOptv1: L-RPT with 128K entries. Block addresses are hashed with a 17-bit bitmask. 
    \item LOptv2: L-RPT with 256K entries. Block addresses are hashed with an 18-bit bitmask. 
    \item LOptv3: L-RPT with 128K entries. Block addresses are hashed using SplitMix32 to generate a 32-bit hash value, from which the lower 17 bits are used as the final index.
    \item LOptv4: L-RPT with 256K entries. Block addresses are hashed using SplitMix32 to generate a 32-bit hash value, from which the lower 18 bits are used as the final index. 
\end{itemize}
We observe that compared to HyDRA (with an unoptimized L-RPT table), these designs show at most 2\% IPC reduction and up to 8\% higher DMR (\emph{mix8} on Config-4), while Configs 5 and 7 show no performance change.}

\subsection{{Performance Evaluation with SHIP-driven bypass with predictor table of similar overhead as the LERN predictor.}}
{In this work, the SHIP predictor is implemented with a 3-bit saturating counter and 4K table entries. To analyze the impact of increasing the predictor table size (comparable to L-RPT size of 320KB) and the counter size, on the accelerator's performance, we modified the SHIP implementation to use an 8-bit saturating counter and a 128K-entry predictor table. In this configuration, the hardware overhead of two SHIP-predictor tables (one for all cores and one for the accelerator) is 256 KB (128 KB each). Figure \ref{fig:shipLarge} demonstrates the performance achieved by HyDRA over FIFO-NB and deadline-agnostic SHIP-driven bypass policies, ARP-CS-AS (4K-entries, 3-bit counter), and ARP-CS-AS-Large (128K-entries, 8-bit counter). We observe that with a larger predictor table, SHIP lowers the bypass rate for processor cores, likely due to lower cross-interference, but does not have a significant impact on the accelerator bypass rate compared to SHIP with a smaller predictor table. We also observe that without deadline awareness, reuse-aware SHIP-driven bypass shows a high DMR of 1, even though the achieved IPC is comparable to HyDRA. We have similar observations for Config-3 and Config-5, with different memory intensity and access patterns, compared to Config-1. Therefore, even with similar overhead, online training might not always be as accurate as offline training for a reuse predictor.} 

We conclude that HyDRA consistently outperforms the SHIP-driven ARP-CS-AS-D policy. It can achieve a higher accelerator BR than SHIP-driven bypass without affecting the DMR. It shows up to 90\% DMR reduction with up to 9\% IPC degradation, and up to 50\% DMR reduction with up to 7.3\% IPC gain over the FIFO-NB policy. 
Therefore, it efficiently manages shared cache resources between the accelerator and the cores to maximize system throughput while meeting the accelerator's deadline. 

\subsection{Parameter Selection}
\label{sec:paramSel}

The parameters of HyDRA are finalized based on experiments across a different set of 30 workload mixes, running on cores and the accelerator configuration Config-3 at 10 IPS.

\subsubsection{Margin High and Low.}
$\mbox{margin}_{\mbox{\scriptsize high}}$ is varied between \{1\%, 2\%, 3\%, 4\%, 5\%, 7\%\} of the required deadline with marginal performance variations.
We observe $\mbox{margin}_{\mbox{\scriptsize high}}$ of 5\% generalizes well across different configurations.
We vary the values of $\mbox{margin}_{\mbox{\scriptsize low}}$ between \{0.5\%, 1\%, 2\%\} and observe that with $\mbox{margin}_{\mbox{\scriptsize high}}$ of 5\%, $\mbox{margin}_{\mbox{\scriptsize low}}$ of 1\% performs best. 

\subsubsection{Miss Rate Threshold.}
We vary $\mbox{MR}_{\mbox{\scriptsize Th}}$ between \{0.1, 0.2, 0.3, 0.5, 0.7, 0.9\}. Higher thresholds do not meet the deadline, while lower ones cause IPC degradation. 
We choose $\mbox{MR}_{\mbox{\scriptsize Th}}$ = 0.3 as it demonstrates high performance across configurations.

\subsubsection{Tolerance Parameters $\alpha, \beta$.}
We vary $\alpha$ across \{0.02, 0.05, 0.1, 0.15, 0.2\} and $\beta$ across \{0.01, 0.02, 0.03, 0.05, 0.07, 0.1\}. We choose $\alpha = 0.1$ and $\beta = 0.05$ as they give comparable speedup and perform well on all configurations.

\subsubsection{Threshold Change Step Size $\delta_A, \delta_B$.}
We vary $\delta_{\mbox{\scriptsize A}}$ across \{0.1, 0.15, 0.2, 0.25, 0.3, 0.5\} and $\delta_{\mbox{\scriptsize B}}$ across \{0.05, 0.1, 0.15\} and observe minimal performance variations. 
We select $\delta_{\mbox{\scriptsize A}}=0.2$ and $\delta_{\mbox{\scriptsize B}}$ = 0.1, which provide the best performance. 

\subsubsection{Epoch Length.}
We conduct experiments with epoch lengths (ET) of  2K, 10K, 20K, 100K, 200K, 400K, and 1M cycles and observe marginal performance variations. Very large ETs reduce the chances of adaptation. Very small ETs yield insufficient data for decision-making. 
Therefore, we choose an epoch length of 200K cycles to minimize the overhead of collecting performance counters every epoch. 

\subsection{Overhead Analysis}
\label{sec:overhead}

The overheads associated with HyDRA are estimated using the Synopsys 22nm technology library and CACTI.

\subsubsection{Latency Overhead.}

HyDRA's decisions do not lie on the performance-critical path, as the APM is invoked before the completion of the previous epoch to estimate the reuse thresholds, and the bypass decisions are evaluated in parallel with the servicing of accesses from the LLC queues. 
 
The LERN methodology is lightweight, and offline training takes less than 4 hours on a single Intel(R) Xeon(R) Gold 5220R CPU @ 2.20GHz for each network shown in Table \ref{tab:hwaconfig}. 
{Online reuse prediction requires loading the layer-wise clustering information into the L-RPT before each layer begins execution.} This is performed during the layer transition time~\cite{FLASH}. 
The L-RPT is direct-mapped and provides a single-cycle lookup.

\subsubsection{Area Overhead.}
{HyDRA is implemented in the cache controller hardware. 
The computations done by APM and the bypass decision module require additional hardware including comparators (nineteen 3-bit, two 32-bit, two 82-bit, and twelve 114-bit), multipliers (fourteen 50-bit and twelve 32-bit), five 32-bit adders, nine 8-bit adders, seven $2\times1$ multiplexers, twenty seven $4\times1$ multiplexers, eight $8\times1$ multiplexers, two 8-bit shift registers, and logic gates (one 2-input AND gate, five 3-input AND gates, two 2-input OR gate, and eleven 2-input XOR gates). 
The L-RPT and the SHIP-RPT occupy 320KB and 3KB of additional space, respectively. The registers for performance counter use an additional 213B.
HyDRA shows an area overhead of 7.1\% compared to the total cache area in 22nm, of which L-RPT contributes 3.9\%, 3.2\% is from additional logic components, and 0.04\% is from the SHIP RPT. 
}

\subsubsection{Energy Overhead.}
{On average, HyDRA shows a 2\% increase and a 8\% decrease in energy consumption compared to FIFO-NB and ARP-CS-AS-D, respectively. 
On average, HyDRA contributes 3.7\% to total energy consumption. }

\section{Conclusion and Future Work}
\label{sec:conclusion}
We proposed and evaluated a deadline- and reuse-aware policy, HyDRA, for selective caching of memory accesses from processor cores and domain-specific accelerators in heterogeneous SoCs, to maximize system throughput while meeting the accelerators' deadlines. 
The state-of-the-art bypass strategies focus solely on improving IPC by using reuse predictors trained on general-purpose applications and do not consider QoS requirements, such as deadlines. 
We showed that such predictors do not achieve optimal performance with accelerators and proposed a new reuse predictor, LERN, to predict reuse behavior for accelerator accesses. 
HyDRA uses this reuse predictor to enable efficient bypass and significantly improve IPC while meeting deadlines. HyDRA is an efficient and effective shared cache space management strategy in heterogeneous SoCs. 

{The current infrastructure models a non-sliced LLC, similar to commercial SoCs such as the NXP LS2088A~\cite{NXP-LS2088A}. In many modern systems, LLCs are distributed across clusters where cores and accelerators share a local slice, and contention within each slice remains relevant under accelerator deadlines. HyDRA can therefore be applied at the cluster level to manage LLC contention, even in distributed-cache systems. As future work, we plan to evaluate HyDRA using the RUBY memory system in \emph{gem5} to model sliced LLC architectures.}

{This work focuses on workloads where memory reuse remains largely stable across inputs. Workloads, such as GNNs, with varying access patterns and reuse across input sets can be investigated as future work.}
This work explored dynamic LLC space management between processor cores and accelerators with static bandwidth partitioning. A natural extension of this work is to study the dynamic LLC bandwidth and space management between them. 
Bypassing strategies are also prone to increased off-chip memory contention and energy consumption. Dynamic space management strategies can also be explored to trade off system performance with off-chip memory energy consumption.

\bibliographystyle{IEEEtran}  
\bibliography{IEEEabrv,references}

\newpage

\begin{IEEEbiography}[{\includegraphics[width=1in,height=1.25in,clip,keepaspectratio]{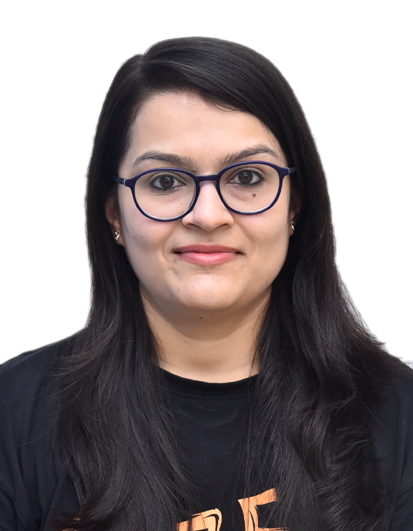}}]{Ayushi Agarwal} received her B.Tech in Electronics and Communication Engineering from Motilal Nehru National Institute of Technology Allahabad, in 2014. She is a research scholar in the ANSK School of Information Technology at the Indian Institute of Technology Delhi. She has previously worked at Qualcomm and as a Research Associate with the GreenIC group at the National University of Singapore. Her research focuses on efficient system integration of domain-specific accelerators into multi-processor SoCs, emphasizing the optimization of shared cache and memory management for improved performance and energy efficiency.
\end{IEEEbiography}

\vskip -2\baselineskip plus -1fil

\begin{IEEEbiography}[{\includegraphics[width=1in,height=1.25in,clip,keepaspectratio]{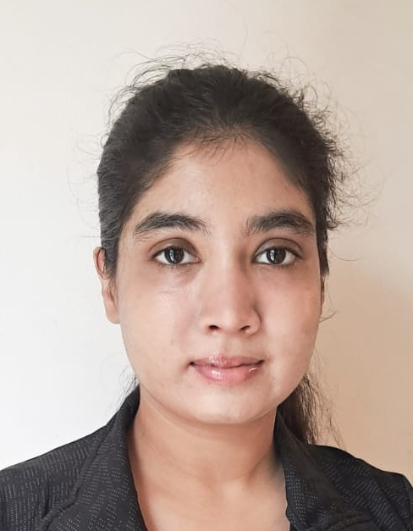}}]{Anannya Mathur} received her Bachelor of Technology degree with a minor in Computer Science and Engineering from the Indian Institute of Technology, Delhi (IIT Delhi) in 2023. She is pursuing a Master’s by Research in the School of Information Technology at IIT Delhi, focusing on Computer Science and Engineering. Her research interests include photonic computing, spiking neural networks, brain-inspired computing, time series prediction, and inverse design algorithms.
\end{IEEEbiography}

\vskip -2\baselineskip plus -1fil

\begin{IEEEbiography}[{\includegraphics[width=1in,height=1.25in,clip,keepaspectratio]{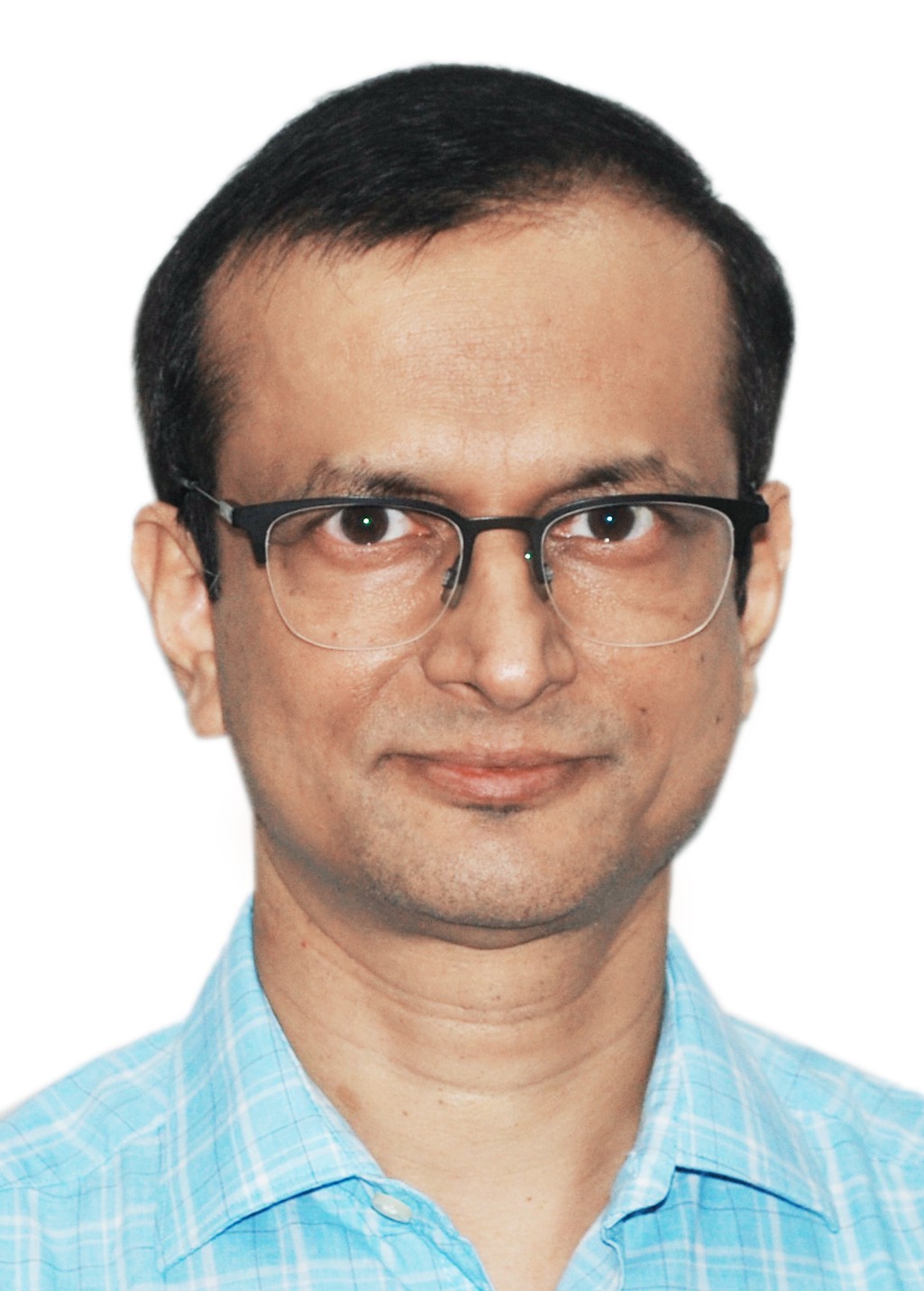}}]{Preeti Ranjan Panda} received his B. Tech. in Computer Science and Engineering from the IIT Madras in 1990 and his M.S. and Ph.D. from the University of California at Irvine in 1995 and 1998, respectively. He is currently a Professor in the Department of Computer Science and Engineering, with a joint appointment at the Khosla School of IT at IIT Delhi. He has previously worked at Texas Instruments and Synopsys, Inc., and as a visiting scholar at Stanford University. He is the author of two books and a recipient of the IBM Faculty Award, the IESA Techno Mentor Award, and the DST Young Scientist Award. He has served as the Editor-in-Chief of IEEE Embedded Systems Letters and on the editorial boards of several journals, including IEEE TCAD, ACM TODAES, and Springer IJPP. He has served as the Technical Program chair of CODES+ISSS and CASES, on the steering committee of ASPDAC, and on the Technical Program Committees of several major conferences, including DAC, ICCAD, and DATE.  
\end{IEEEbiography}

\end{document}